\documentclass[twocolumn]{aa} 
\usepackage{graphicx}
\usepackage{txfonts}
\usepackage{natbib}
\bibpunct{(}{)}{;}{a}{}{,} 
\begin{document}
\title{Structural properties of disk galaxies.}

\subtitle{II. Intrinsic shape of bulges}

\author{J. M\'endez-Abreu\inst{1,2}
        \and
        E. Simonneau\inst{3}
        \and
        J. A. L. Aguerri\inst{1,2}
	\and
	E. M. Corsini\inst{4}
	}

\institute{Instituto Astrof\'\i sico de
           Canarias, Calle V\'ia L\'actea s/n, E-38200 La Laguna, Spain
         \and Departamento de Astrof\'\i sica, Universidad de La Laguna, 
              E-38205 La Laguna, Tenerife, Spain \\
         \email{jairo@iac.es;jalfonso@iac.es} 
         \and Institut d'Astrophysique de Paris, 
              C.N.R.S.-U.P.M.C., 98bis Boul. Arago, F-75014 Paris, France
         \and Dipartimento di Astronomia, Universit\`a di Padova, 
              vicolo dell'Osservatorio~3, I-35122 Padova, Italy\\ 
         \email{enricomaria.corsini@unipd.it} 
         }

   \date{Received September 15, 1996; accepted March 16, 1997}

\abstract 
{
Knowledge of  the intrinsic  shape of galaxy  components is  a crucial
piece of  information to  constrain phenomena driving  their formation
and evolution.}
{
The  structural  parameters  of  a  magnitude-limited  sample  of  148
unbarred S0--Sb  galaxies were analyzed to derive  the intrinsic shape
of their bulges.}
{ 
We  developed a new  method to  derive the  intrinsic shape  of bulges
based  on  the  geometrical  relationships between  the  apparent  and
intrinsic  shapes of  bulges  and  disks.
The  equatorial ellipticity  and intrinsic  flattening of  bulges were
obtained from the length of  the apparent major and minor semi-axes of
the bulge,  twist angle between the  apparent major axis  of the bulge
and the galaxy line of nodes, and galaxy inclination.}
{
We found that the intrinsic  shape is well constrained for a subsample
of 115 bulges with favorable viewing angles  .
A large  fraction of  them is characterized  by an  elliptical section
($B/A<0.9$).  This fraction  is $33\%$,  $55\%$, and  $43\%$  if using
their maximum, mean, or median equatorial ellipticity, respectively.
Most are  flattened along their polar axis  ($C<(A+B)/2$).
The distribution  of triaxiality is strongly  bimodal. This bimodality
is driven by  bulges with S\'ersic index $n >  2$, or equivalently, by
the bulges of galaxies with a bulge-to-total ratio $B/T > 0.3$.
Bulges  with $n  \leq 2$  and  with $B/T  \leq 0.3$  follow a  similar
distribution, which is different from that  of bulges with $n > 2$ and
with $B/T > 0.3$.
In particular,  bulges with  $n\leq2$ and with  $B/T \leq 0.3$  show a
larger  fraction  of  oblate  axisymmetric  (or  nearly  axisymmetric)
bulges,  a smaller  fraction  of triaxial  bulges,  and fewer  prolate
axisymmetric (or  nearly axisymmetric)  bulges with respect  to bulges
with $n > 2$ and with $B/T > 0.3$, respectively.
}
{ 
According  to  predictions  of  the  numerical  simulations  of  bulge
formation,  bulges with  $n \leq  2$, which  show a  high  fraction of
oblate axisymmetric (or nearly axisymmetric) shapes and have $B/T \leq
0.3$, could be  the result of dissipational minor  mergers. Both major
dissipational  and  dissipationless mergers  seem  to  be required  to
explain the variety of shapes found for bulges with $n > 2$ and $B/T >
0.3$.}


\maketitle
%
\section{Introduction}
\label{sec:intro}

The halos  of cold dark  matter assembled in  cosmological simulations
appear   to   be    strongly   triaxial   \citep[see][and   references
  therein]{allgood06}.  Their  intrinsic shape is  characterized by an
intermediate-to-long axis  ratio $B/A$ and a  short-to-long axis ratio
$C/A$ which  can vary as  a function of  radius. On the  contrary, the
halo   shape   inferred   from   observations   of   the   Milky   Way
\citep{ollingmerrifield00,  ibata01,  johnston05}   and  a  number  of
individual    nearby   galaxies    \citep{merrifield04}    is   nearly
axisymmetric.
The study of  the intrinsic shape of the  luminous galactic components
may serve to  constrain the halo shape, which is  related to the final
morphology  of the  galaxy and  depends on  the phenomena  driving its
formation  and  evolution  \citep[e.g.,][]{heller07}.   The  intrinsic
shapes of  elliptical galaxies and  disks were extensively  studied in
the  past, whereas  bulges appear  to be  less studied,  even  if they
account for  about $25\%$  of the stellar  mass of the  local universe
\citep{driver07}.

\subsection{Intrinsic shape of elliptical galaxies}

The first attempt to derive the intrinsic shape of elliptical galaxies
was  done by  \citet{hubble26}.  The  distribution of  their intrinsic
flattenings  was obtained  from the  observed ellipticities  under the
assumption  that elliptical  galaxies  were oblate  ellipsoids with  a
random orientation with respect to the line of sight.
Early  studies  considered  elliptical  galaxies  to  be  axisymmetric
systems.  Oblateness and prolateness were assumed by \citet{sandage70}
and  \citet{binney78}, respectively to  reproduce the  distribution of
observed  ellipticities of  the Reference  Catalog of  Bright Galaxies
\citep[][hereafter RC1]{devaucouleurs64}.

Afterwards, a number of kinematic  and photometric findings led to the
suggestion  that there are  also elliptical  galaxies with  a triaxial
shape. In fact, the low ratio between rotational velocity and velocity
dispersion \citep{bertolacapaccioli75, illingworth77}, the twisting in
the isophotes \citep{carter78, bertolagalletta79, galletta80}, and the
rotation measured along the minor axis \citep{schechtergunn79} of some
elliptical galaxies  could not be  explained in terms  of axisymmetric
ellipsoids.    As  a   consequence,   \citet{benacchiogalletta80}  and
\citet{binneydevaucouleurs81} showed that the distribution of observed
ellipticities could be satisfactorily accounted for also in terms of a
distribution of triaxial ellipsoids.
Similar    conclusions    were    reached   by    \citet{fasanovio91},
\citet{lambas92},  \citet{ryden92,   ryden96},  and  \citet{fasano95}.
However,  different  galaxy   samples  and  different  assumptions  on
triaxiality  resulted in  different distributions  of  intrinsic axial
ratios.
In addition, not  all the elliptical galaxies have  the same intrinsic
shape. In fact,  \citet{tremblaymerritt96} found that the distribution
of the observed ellipticities of galaxies brighter than $M_B\simeq-20$
is different than that of the less luminous ones. In particular, there
is  a  relative lack  of  highly-flattened  bright ellipticals.   This
reflects   a   difference  in   the   shape   of  low-luminosity   and
high-luminosity   ellipticals:  fainter  ellipticals   are  moderately
flattened  and  oblate, while  brighter  ellipticals  are rounder  and
triaxial.  Recently, \citet{fasano10}  found  also that  even if  both
normal ellipticals and brightest  cluster galaxies (BCG) are triaxial,
BCGs tend  to have  a more  prolate shape, and  that this  tendency to
prolateness  is mainly  driven by  the  cD galaxies  present in  their
sample of BCGs.
These kinds of statistical analyses benefit from large galaxy samples,
such     as     those      studied     by     \citet{kimmyi07}     and
\citet{padillastrauss08}. They analyzed  the observed ellipticities of
early-type  galaxies in  Sloan Digital  Sky  Survey \citep{adelman06}.
Furthermore, these large datasets allowed them to study the dependence
of  the  intrinsic shape  on  other  galaxy  properties, such  as  the
luminosity, color, physical size, and environment.

Determining  the distribution  of  the intrinsic  shape of  elliptical
galaxies  is  also possible  by  combining  photometric and  kinematic
information   \citep{binney85,  franx91}.    However,   the  resulting
distribution of  intrinsic flattenings, equatorial  ellipticities, and
intrinsic misalignments between the angular momentum and the intrinsic
short  axis can  not be  derived  uniquely. Only  two observables  are
indeed available,  the distribution of observed  ellipticities and the
distribution of kinematic  misalignments between the photometric minor
axis and the kinematic  rotation axis.  Therefore, further assumptions
about the  intrinsic shape and  direction of the angular  momentum are
needed to simplify the problem.  In addition, this approach requires a
large  sample of  galaxies  for which  the  kinematic misalignment  is
accurately measured.  But, to  date this information is available only
for a few tens of galaxies \citep{franx91}.

Many individual galaxies have  been investigated by detailed dynamical
modeling  of  the kinematics  of  gas,  stars,  and planetary  nebulae
\citep[e.g.,][]{tenjes93,  statler94a,  statler94b, mathieudejonghe99,
  gerhard01, gebhardt03, cappellari07,thomas07, delorenzi09}.
Recently, \citet{vandenboschvandeven09} have investigated how well the
intrinsic  shape of elliptical  galaxies can  be recovered  by fitting
realistic  triaxial  dynamical  models  to simulated  photometric  and
kinematic observations.  The recovery  based on orbit-based models and
state-of-the-art  data   is  degenerate  for   round  or  not-rotating
galaxies. The intrinsic flattening of oblate ellipsoids is almost only
constrained  by  photometry.   The   shape  of  triaxial  galaxies  is
accurately  determined  when   additional  photometric  and  kinematic
complexity,  such  as  the  presence  of  an  isophotal  twist  and  a
kinematically decoupled core, is observed.
Finally,  the  intrinsic shape  of  individual  galaxies  can be  also
constrained  from  the observed  ellipticity  and  isophotal twist  by
assuming   the  intrinsic   density   distribution  \citep{williams81,
  chakraborty08}.

\subsection{Intrinsic shape of disk galaxies}

Although  the  disks  of  lenticular  and spiral  galaxies  are  often
considered to  be infinitesimally  thin and perfectly  circular, their
intrinsic   shape  is  better   approximated  by   flattened  triaxial
ellipsoids.

The disk  thickness can be directly determined  from edge-on galaxies.
It depends both  on the wavelength at which disks  are observed and on
galaxy morphological type. Indeed, galactic disks become thicker
at  longer wavelengths  \citep{dalcantonbernstein02,  mitronova04} and
late-type   spirals  have  thinner   disks  than   early-type  spirals
\citep{bottinelli83, guthrie92}.

Determining the distribution of  both the thickness and ellipticity of
disks is  possible by  a statistical analysis  of the  distribution of
apparent axial ratios of randomly oriented spiral galaxies.
\citet{sandage70} analyzed the spiral galaxies listed in the RC1. They
concluded that disks are circular  with a mean flattening $\langle C/A
\rangle = 0.25$.
However, the  lack of  nearly circular spiral  galaxies ($B/A\simeq1$)
rules  out that disks  have a  perfectly axisymmetric  shape.  Indeed,
\citet{binggeli80},          \citet{benacchiogalletta80},          and
\citet{binneydevaucouleurs81}   showed   that   disks   are   slightly
elliptical with a mean ellipticity $\langle 1-B/A \rangle = 0.1$.
These early findings were based on the analysis of photographic plates
of a few hundreds of galaxies. Later, they were confirmed by measuring
ellipticities  of  several thousands  of  objects  in  CCD images  and
digital  scans of plates  obtained in  wide-field surveys.   The large
number of objects  allows the constraining of the  distribution of the
intrinsic equatorial ellipticity, which  is well fitted by a one-sided
Gaussian centered on $1-B/A=0$  with a standard deviation ranging from
0.1 to 0.2  and a mean of 0.1  \citep{lambas92, fasano93, alamryden02,
  ryden04}.
Like  the  flattening,  the   intrinsic  ellipticity  depends  on  the
morphological type  and wavelength.   The disks of  early-type spirals
are more elliptical  than those of late-type spirals  and their median
ellipticity   increases  with  observed   wavelength  \citep{ryden06}.
Furthermore, luminous spiral galaxies tend to have thicker and rounder
disks        than         low-luminosity        spiral        galaxies
\citep{padillastrauss08}. Different  mechanisms have been  proposed to
account  for disk thickening,  including the  scattering of  stars off
giant   molecular  clouds  \citep{spitzerschwarzschild51,villumsen85},
transient      density     waves      of      the     spiral      arms
\citep{barbaniswoltjer67,carlbergsellwood85},  and minor  mergers with
satellite galaxies \citep[e.g.,][]{quinn93,walker96}.

The  study of  the intrinsic  shape of  bulges  presents similarities,
advantages, and drawbacks with respect to that of elliptical galaxies.
For  bulges, the  problem  is  complicated by  the  presence of  other
luminous  components  and  requires   the  isolation  of  their  light
distribution.   This  can  be  achieved by  performing  a  photometric
decomposition  of  the  galaxy surface-brightness  distribution.   The
galaxy light is usually modeled as the sum of the contributions of the
different galactic  components, i.e.,  bulge and disk,  and eventually
lenses, bars,  spiral arms, and  rings \citep[][]{prieto01,aguerri05}.
A number  of two-dimensional parametric  decomposition techniques have
been   developed   in  the   last   several   years   with  this   aim
\citep[e.g.,][]{simard98,     khosroshahi00,     peng02,    desouza04,
  laurikainen05, pignatelli06, mendezabreu08a}.
On the  other hand, the presence  of the galactic disk  allows for the
accurate  constraining  of the  inclination  of  the  bulge under  the
assumption that  the two components  share the same polar  axis (i.e.,
the equatorial plane of the disk coincides with that of the bulge).

As elliptical galaxies, bulges  are diverse and heterogeneous objects.
Big  bulges  of lenticulars  and  early-type  spirals  are similar  to
low-luminosity elliptical  galaxies. On the contrary,  small bulges of
late-type  spirals are  reminiscent  of disks  \citep[see the  reviews
  by][]{kormendy93, wyse97, kormendykennicutt04}.  Some of them have a
quite  complex  structure and  host  nuclear  rings \citep[see][for  a
  compilation]{buta95,comeron10},   inner   bars   \citep[see][for   a
  list]{erwin04},  and embedded  disks \citep[e.g.,][]{scorzabender95,
  vandenbosch98,pizzella02}.
Although the kinematical properties  of many bulges are well described
by  dynamical  models of  oblate  ellipsoids  which  are flattened  by
rotation  with little  or no  anisotropy \citep{kormendyillingworth82,
  daviesillingworth83,     fillmore86,     corsini99,    pignatelli01,
  cappellari06},    the    twisting    of    the    bulge    isophotes
\citep{lindblad56,  zaritskylo86}  and  the misalignment  between  the
major   axes  of   the  bulge   and   disk  \citep{bertola91,varela96,
  mendezabreu08a} observed in several galaxies are not possible if the
bulge and disk are both axisymmetric.  These features were interpreted
as the signature of bulge triaxiality.
This idea  is supported  by the presence  of non-circular  gas motions
\citep[e.g.,][]{gerhardvietri86,   bertola89,   gerhard89,   berman01,
  falconbarroso06,  pizzella08}  and  a  velocity gradient  along  the
galaxy minor axis \citep[e.g.,][]{corsini03, coccato04, coccato05}.

Perfect  axisymmetry is  also ruled  out when  the intrinsic  shape of
bulges is  determined by statistical analyses based  on their observed
ellipticities.
\citet{bertola91} measured the  bulge ellipticity and the misalignment
between  the  major   axes  of  the  bulge  and   disk  in  32  S0--Sb
galaxies. They  found that these  bulges are triaxial with  mean axial
ratios $\langle B/A \rangle=0.86$ and $\langle C/A \rangle=0.65$.
$\langle  B/A\rangle =  0.79$ for  the  bulges of  35 early-type  disk
galaxies  and  $\langle  B/A\rangle  =  0.71$ for  the  bulges  of  35
late-type spirals studied by \citet{fathipeletier03}. They derived the
equatorial ellipticity by analyzing the deprojected ellipticity of the
ellipses fitting the galaxy isophotes within the bulge radius.
None of the  21 disk galaxies with morphological  types between S0 and
Sab  studied  by   \citet{noordermeervanderhulst07}  harbors  a  truly
spherical  bulge. A  mean  flattening $\langle  C/A \rangle=0.55$  was
obtained  under the assumption  of bulge  oblateness by  comparing the
isophotal ellipticity  in the bulge-dominated region  to that measured
in the disk-dominated region.
\citet{mosenkov10} obtained a median  value of the flattening $\langle
C/A \rangle=0.63$  for a  sample of both  early and  late-type edge-on
galaxies  in the  near  infrared.  They found  also  that bulges  with
S\'ersic  index $n<2$  can be  described by  triaxial,  nearly prolate
bulges that are  seen from different projections, while  $n>$ 2 bulges
are better represented by oblate spheroids with moderate flattening.

In  \citet[][hereafter   Paper  I]{mendezabreu08a}  we   measured  the
structural parameters  of a  magnitude-limited sample of  148 unbarred
early-to-intermediate  spiral  galaxies   by  a  detailed  photometric
decomposition     of     their    near-infrared     surface-brightness
distribution. The probability distribution function (PDF) of the bulge
equatorial ellipticity was derived  from the distributions of observed
ellipticities of bulges and misalignments between bulges and disks. We
proved  that about  $80\%$ of  the sample  bulges are  not  oblate but
triaxial  ellipsoids with  a mean  axial ratio  $\langle  B/A\rangle =
0.85$.  The  PDF is not  significantly dependent on  morphology, light
concentration,  or  luminosity  and  is independent  on  the  possible
presence of nuclear bars.  This is by far the largest sample of bulges
studied with this aim.

In this paper, we introduce a new method to derive the intrinsic shape
of  bulges  under the  assumption  of  triaxiality.  This  statistical
analysis is  based upon the analytical relations  between the observed
and intrinsic shapes  of bulges and their surrounding  disks and it is
applied to the galaxy sample described in Paper I.
The method was  conceived to be completely independent  of the studied
class of objects,  and it can be applied  whenever triaxial ellipsoids
embedded in (or embedding) an axisymmetric component are considered.

The structure of the paper is as follows. The basic description of the
geometry  of the  problem  and  main definitions  are  given in  Sect.
\ref{sec:basic}.    The  statistical   analysis   of  the   equatorial
ellipticity and  intrinsic flattening of bulges is  presented in Sect.
\ref{sec:ellip} and \ref{sec:flat}, respectively.  The intrinsic shape
of bulges  is discussed in  Sect.  \ref{sec:3d}.  The  conclusions are
presented in Sect. \ref{sec:conclu}.

\section{Basic geometrical considerations}
\label{sec:basic}

As in  Paper I, we assume that  the bulge is a  triaxial ellipsoid and
the disk  is circular and lies  in the equatorial plane  of the bulge.
Bulge and  disk share the same  center and polar  axis. Therefore, the
inclination of the  polar axis (i.e., the galaxy  inclination) and the
position angle of  the line of nodes (i.e., the  position angle of the
galaxy major axis) are  directly derived from the observed ellipticity
and orientation of the disk, respectively.

We  already introduced in  Paper I  the basic  geometrical definitions
about  the  triaxial  ellipsoidal  bulge  and its  deprojection  as  a
function of the main parameters  which describe the problem, i.e., the
ellipticity $e$ of the projected ellipse, twist angle $\delta$ between
its major axis and the line of nodes, galaxy inclination $\theta$, and
orientation $\phi$ of the equatorial axes of the bulge with respect to
the line  of nodes.  However, for  the sake of clarity  we will review
again these concepts in this section together with the new definitions
needed to perform our statistical approach.

\subsection{Direct problem: from ellipsoids to ellipses}

Let ($x,  y, z$) be the  Cartesian coordinates with the  origin in the
galaxy  center,  the  $x-$axis   and  $y-$axis  corresponding  to  the
principal equatorial axes of the bulge, and the $z-$axis corresponding
to the  polar axis.   As the equatorial  plane of the  bulge coincides
with the equatorial plane of the  disk, the $z-$axis is also the polar
axis  of the  disk.   If $A$,  $B$, and  $C$  are the  lengths of  the
ellipsoid semi-axes,  the corresponding equation  of the bulge  in its
own reference system is given by

\begin{equation} 
\frac{x^2}{A^2} + \frac{y^2}{B^2} + \frac{z^2}{C^2} = 1. 
\label{eqn:ellipsoid} 
\end{equation} 
%
It is worth  noting that we do not  assume that $A \geq B  \geq C$, as
usually done in the literature.

Let  $(x',y',z')$ be  now the  Cartesian coordinates  of  the observer
system. It  has its origin in  the galaxy center,  the polar $z'-$axis
along  the line of  sight (LOS)  and pointing  toward the  galaxy. The
plane of the sky lies on the $(x',y')$ plane.
 
\begin{figure*}[!ht]
  \centering
  \includegraphics[width=\textwidth]{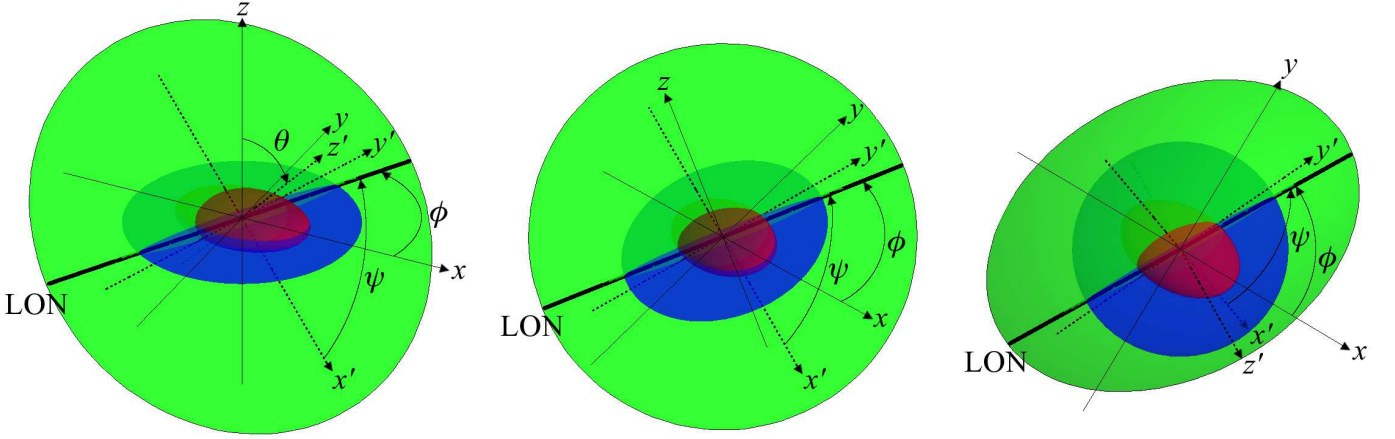}
  \caption{Schematic three-dimensional view of the ellipsoid geometry.
    The bulge ellipsoid,  the disk plane, and the  sky plane are shown
    in red,  blue, and green, respectively.  The  reference systems of
    the ellipsoid and  of the observer as well as  the LON are plotted
    with thin solid lines, thin  dashed lines, and a thick solid line,
    respectively.  The  bulge  ellipsoid  is  shown as  seen  from  an
    arbitrary  viewing  angle (left  panel),  along  the LOS  (central
    panel),  and  along the  polar  axis  (i.e.,  the $z-$axis;  right
    panel).}
  \label{fig:geom}
  \end{figure*}

The projection  of the  disk onto  the sky plane  is an  ellipse whose
major axis is the line  of nodes (LON), i.e., the intersection between
the galactic and  sky planes. The angle $\theta$  between the $z-$axis
and  $z'-$axis  corresponds  to  the  inclination of  the  galaxy  and
therefore   of   the  bulge   ellipsoid;   it   can   be  derived   as
$\theta=\arccos{(d/c)}$  from  the  length  $c$  and $d$  of  the  two
semi-axes of the projected ellipse of the disk.
We defined $\phi$ ($0 \leq \phi \leq \pi /2$) as the angle between the
$x$-axis and the LON on the equatorial plane of the bulge $(x,y)$.
Finally, we  also defined $\psi$  ($0 \leq \psi  \leq \pi /2$)  as the
angle  between   the  $x'$-axis   and  the  LON   on  the   sky  plane
$(x',y')$. The three angles $\theta$, $\phi$, and $\psi$ are the usual
Euler  angles  and  relate  the  reference  system  $(x,y,z)$  of  the
ellipsoid with  that $(x',y',z')$  of the observer  by means  of three
rotations (see  Fig. \ref{fig:geom}). Indeed, because  of the location
of  the LON  is  known, we  can  choose the  $x'-$axis  along it,  and
consequently it holds that $\psi=0$.
By  applying these  two rotations  to Eq.   \ref{eqn:ellipsoid}  it is
possible  to derive  the  equation  of the  ellipsoidal  bulge in  the
reference  system of  the observer,  as well  as the  equation  of the
ellipse   corresponding   to  its   projection   on   the  sky   plane
\citep{simonneau98}.
Now,  if we  identify the  latter with  the ellipse  projected  by the
observed ellipsoidal bulge, we can  determine the position of its axes
of symmetry $x_{\rm e}$ and $y_{\rm e}$ and the lengths $a$ and $b$ of
the  corresponding semi-axes.  The  $x_{\rm e}-$  axis forms  an angle
$\delta$  with the  LON  corresponding  to the  $x'-$axis  of the  sky
plane. We always choose $0 \le  \delta \le \pi/2$ such that $a$ can be
either the  major or the minor  semi-axis.  If $a$  corresponds to the
major semi-axis then $b$ is the  length of the minor semi-axis. If $a$
corresponds to the minor semi-axis then $b$ is the length of the major
semi-axis.  Later in this paper,  when we will present our statistical
analysis  we will  find that  this riddle  is solved  because  the two
possibilities coincide, and one is the mirror image of the other.

From the previous considerations \citep[see][for details]{simonneau98}
we have that the equations relating the length of the semi-axes of the
projected ellipse  with the length  of the semi-axes of  the intrinsic
ellipsoid are given by

\begin{scriptsize} 
\begin{eqnarray} 
a^{2}b^{2} &=& A^{2}C^{2}\sin^{2}\theta \cos^{2}\phi + 
  B^{2}C^{2}\sin^{2}\theta \sin^{2}\phi + A^{2}B^{2}\cos^{2}\theta , 
\label{eqn:ab}\\ 
a^{2}+b^{2}&=&A^{2}(\cos^{2}\phi + \cos^{2}\theta \sin^{2}\phi) +  
  B^{2}(\sin^{2}\phi + \cos^{2}\theta \cos^{2}\phi) + C^{2}\sin^{2}\theta ,  
\label{eqn:a+b}\\ 
\tan2\delta&=&\frac{(B^{2}-A^{2})\cos\theta\sin2\phi}{A^{2}(\cos^{2} 
  \theta\sin^{2}\phi -\cos^{2}\phi) +  
  B^{2}(\cos^{2}\theta\cos^{2}\phi - \sin^{2}\phi) + C^{2}\sin^{2}\theta}. 
\label{eqn:tandelta} 
\end{eqnarray} 
\end{scriptsize} 
 
If the ellipsoidal  bulge is not circular in  the equatorial plane ($A
\ne  B$) then it  is possible  to observe  a twist  ($\delta\ne0$; see
Eq. \ref{eqn:tandelta}) between the  axes of the projected ellipses of
the bulge and disk.

\subsection{Inverse problem: from ellipses to ellipsoids} 
\label{sec:deprojection} 

We  will focus now  our attention  on the  inverse problem,  i.e., the
problem   of   deprojection.    Following  \citet{simonneau98},   from
Eqs. \ref{eqn:ab}, \ref{eqn:a+b},  and \ref{eqn:tandelta}, we are able
to express the length of the  bulge semi-axes ($A$, $B$, and $C$) as a
function of the length of the semi-axes of the projected ellipse ($a$,
$b$) and the twist angle ($\delta$).

For the sake  of clarity, we rewrite here  the corresponding equations
but in a different way with respect to Paper I.  First, we define

\begin{equation} 
K^2 = \frac{a^2+b^2}{2}\left[ 1 + e\,\cos 2\delta \right], 
\end{equation} 
%
where 
\begin{equation} 
e=\frac{a^2  - b^2}{a^2 + b^2}  \qquad -1 \le e \le 1,
\label{eqn:e}
\end{equation} 
%
is,  in some  sense,  a measure  of  the ellipticity  of the  observed
ellipse. Therefore, $K^2$ is a positive measurable quantity.

From  Eqs.  \ref{eqn:ab},  \ref{eqn:a+b},  and  \ref{eqn:tandelta}  it
results

\begin{equation} 
K^2 = \frac{A^2+B^2}{2}\left[ 1 + E\,\cos 2\phi \right],
\label{eqn:K}
\end{equation} 
%
where
\begin{equation} 
E=\frac{A^2 - B^2}{A^2 + B^2} \qquad -1 \le E \le 1,
\label{eqn:E}
\end{equation} 
%
measures the intrinsic equatorial ellipticity of the bulge.

With this notation we can rewrite the equations for the semi-axes of
the bulge in the form
 
\begin{small} 
\begin{eqnarray} 
A^2 & = & K^2\left( 1 + \frac{e\,\sin2\delta}{1 + e\,\cos2\delta}\frac{\tan\phi}{\cos\theta}\right) ,  \label{eqn:A} \\ 
B^2 & = & K^2\left( 1 - \frac{e\,\sin2\delta}{1 + e\,\cos2\delta}\frac{\cot{\phi}}{\cos\theta}\right) , \label{eqn:B} \\ 
C^2 & = & K^2\left( 1 - \frac{2\,e\,\cos2\delta}{\sin^2\theta\left(1 + e\,\cos2\delta\right)} + \frac{2\,e\,\cos\theta\,\sin2\delta}{\sin^2\theta\left(1 + e\,\cos2\delta\right)} \cot^2{\phi} \right) \label{eqn:C}. 
\end{eqnarray} 
\end{small} 

The values  of $a, b, \delta,$  and $\theta$ can  be directly obtained
from observations.  Unfortunately, the relation  between the intrinsic
and projected  variables also depends  on the spatial position  of the
bulge  (i.e., on  the  $\phi$  angle), which  is  actually the  unique
unknown of our problem. Indeed,  this will constitute the basis of our
statistical analysis.

\subsection{Characteristic angles} 
\label{sec:angles} 

There  are physical  constraints which  limit the  possible  values of
$\phi$,  such as the  positive length  of the  three semi-axes  of the
ellipsoid \citep{simonneau98}.
Therefore, we  define some  characteristic angles which  constrain the
range  of $\phi$.   Two  different possibilities  must  be taken  into
account for any value of the observed variables $a$, $b$, $\delta$ and
$\theta$.

The first  case corresponds to $a>b$.  It implies that  $e>0$ from Eq.
\ref{eqn:e} and $A>B$ from Eqs.  \ref{eqn:A} and \ref{eqn:B}.  For any
value of $\phi$,  $A^2>K^2$ and $K^2$ is always  positive according to
Eq.  \ref{eqn:K}.  On the  other hand, $B^2$  and $C^2$ can  be either
positive or  negative depending  on the value  of $\phi$  according to
Eqs.   \ref{eqn:B}  and \ref{eqn:C},  respectively.   This limits  the
range  of the values  of $\phi$.   $B^2$ is  positive only  for $\phi>
\phi_B$.    The  angle   $\phi_B$  is   defined  by   $B^2  =   0$  in
Eq. \ref{eqn:B} as

\begin{equation}
\tan{\phi_B}= \frac{e\,\sin{2\delta}}{\cos{\theta}\,\left( 1+e\,\cos{2\delta}\right)}.
\label{eqn:B=0}
\end{equation}

Likewise,  $C^2$ is  positive only  for values  of  $\phi<\phi_C$. The
angle $\phi_C$ is defined by $C^2=0$ in Eq. \ref{eqn:C} as

\begin{equation}
\tan{2\phi_C}= \frac{2\,e\,\sin{2\delta}\,\cos{\theta}}{e\,\cos{2\delta}\,\left(1+\cos^2{\theta}\right)-\sin^2{\theta}}.
\label{eqn:C=0}
\end{equation}

Thus, if  $a>b$ then  the values of  $\phi$ can  only be in  the range
$\phi_B \le \phi \le \phi_C$.

The  second case  corresponds to  $a<b$.  It  implies that  $e<0$ (Eq.
\ref{eqn:e}) and  $A<B$ (Eqs.  \ref{eqn:A} and  \ref{eqn:B}).  For any
value of $\phi$,  $B^2>K^2$ and $K^2$ is always  positive according to
Eq.   \ref{eqn:K}. But,  $A^2$ and  $C^2$  can be  either positive  or
negative  depending   on  the  value  of  $\phi$   according  to  Eqs.
\ref{eqn:A} and  \ref{eqn:C}, respectively.  This limits  the range of
the values  of $\phi$.   $A^2$ is positive  only for $\phi  < \phi_A$.
The angle $\phi_A$ is defined by $A^2 = 0$ in Eq. \ref{eqn:A} as

\begin{equation}
\tan{\phi_A}= -\frac{\cos{\theta}\left(1+e\,\cos{2\delta}\right)}{e\,\sin{2\delta}}.
\label{eqn:A=0}
\end{equation}

Likewise,  $C^2$ is  positive only  for values  of  $\phi>\phi_C$. The
angle $\phi_C$  is given in  Eq. \ref{eqn:C=0}.  Thus, if  $a<b$, then
the  values $\phi$  can only  be  in the  range $\phi_C  \le \phi  \le
\phi_A$.

However, the  problem is  symmetric: the second  case, when  the first
semi-axis of  the observed ellipse  (which is measured  clockwise from
the LON)  corresponds to the minor  axis (i.e., $a<b$),  is the mirror
situation of the first case,  when the first measured semi-axis of the
observed ellipse corresponds  to the major axis (i.e.,  $a>b$). In the
second  case, if we  assume the  angle $\pi/2  -\delta$ to  define the
position  of the  major semi-axis  $a$  of the  observed ellipse  with
respect to the  LON in the sky plane, and $\pi/2  -\phi$ to define the
position of the  major semi-axis $A$ of the  equatorial ellipse of the
bulge with respect  to the LON in the bulge  equatorial plane, then we
can always consider  $a>b$ and $A>B$.  Therefore, $e\ge  0$ and $E \ge
0$ always.  This means that  we have the same mathematical description
in both cases: the possible values  of $\phi$ are $\phi_B \le \phi \le
\phi_C$ with $\phi_B$ and  $\phi_C$ defined by Eqs.  \ref{eqn:B=0} and
\ref{eqn:C=0},  respectively.    Furthermore,  we  can   rewrite  Eqs.
\ref{eqn:A}, \ref{eqn:B}, and \ref{eqn:C}

\begin{small} 
\begin{eqnarray} 
A^2 & = & K^2\left[1 + \tan{\phi_B}\,\tan{\phi}\right] ,  \label{eqn:A_K} \\ 
B^2 & = & K^2\left[1 - \frac{\tan{\phi_B}}{\tan{\phi}}\right] , \label{eqn:B_K} \\ 
C^2 & = & K^2\,2\,\tan{\phi_B}\frac{\cos^2\theta}{\sin^2\theta}\left[\cot{2\,\phi}-\cot{2\,\phi_C}\right] \label{eqn:C_K}. 
\end{eqnarray} 
\end{small} 
%
where $\phi_B$  and $\phi_C$ are given  as a function  of the observed
variables  $a$, $b$,  $\delta$,  and $\theta$,  i.e.,  they are  known
functions for each observed bulge.

We can  always consider  $A>B$ as explained  before.  But, we  are not
imposing   any   constraint  on   the   length   $C$   of  the   polar
semi-axis. According  to this  definition oblate and  prolate triaxial
ellipsoids  do not necessarily  have an  axisymmetric shape.   We will
define a  triaxial ellipsoid  as completely oblate  if $C$  is smaller
than both  $A$ and $B$ (i.e., the  polar axis is the  shortest axis of
the ellipsoid).  We define  a triaxial ellipsoid as completely prolate
if $C$ is greater  than both $A$ and $B$ (i.e., the  polar axis is the
longest axis of the ellipsoid).  If the polar axis is the intermediate
axis we have either a partially oblate or a partially prolate triaxial
ellipsoid.  A further detailed description  of all these cases will be
given at the end of this section.

From Eq.   \ref{eqn:B_K} we  obtain that the  semi-axis length  $B$ is
zero for $\phi=\phi_B$ and it increases when $\phi$ goes from $\phi_B$
to $\phi_C$.  The  semi-axis length $C$ is zero  for $\phi=\phi_C$ and
decreases when  $\phi$ goes  from $\phi_B$ to  $\phi_C$.  There  is an
intermediate  value $\phi_{BC}$  for  which $B^2=C^2$.  This angle  is
given by

\begin{equation}
\tan{\phi_{BC}}= \frac{\tan{\delta}}{\cos{\theta}}.
\end{equation}

For $\phi_{BC}<\phi<\phi_C$,  $C^2<B^2$ and  both of them  are smaller
than  $A^2$.   This   implies  that  in  this  range   of  $\phi$  the
corresponding triaxial ellipsoid is completely oblate.

On the other hand, $B^2<A^2$  for all possible values of $\phi$.  This
is not the case for $C^2$, because it increases when $\phi$ decreases.
Thus, we can define a  new angle $\phi_{AC}$ for which $C^2=A^2$. This
angle is given by

\begin{equation}
\tan{\phi_{AC}}= \cos{\theta}\tan{\delta}.
\end{equation}

For  $\phi<\phi_{AC}$,  $C^2>A^2>B^2$.   Therefore, the  corresponding
triaxial ellipsoid  is completely prolate.  It is  important to notice
here   that    this   case   is   physically    possible   only   when
$\phi_{AC}>\phi_B$, and the  values of $\phi$ are within  the range of
possible values $\phi_{B} \le \phi \le \phi_C$.
Therefore, we conclude that for  any observed bulge (i.e., for any set
of  measured   values  of  $a$,  $b$,  $\delta$,   and  $\theta$)  the
corresponding  triaxial ellipsoid could  be always  completely oblate,
while we are not sure that it could be prolate.

We define the  quadratic mean radius of the  equatorial ellipse of the
bulge   in   order   to   extensively  discuss   all   the   different
possibilities.

\begin{equation}
R^2 = \frac{A^2+B^2}{2} = K^2\tan{\phi_B}\left[\cot{\phi_B} - \cot{2\,\phi}\right],
\label{eqn:R}
\end{equation}
%
which depends only on the unknown position $\phi$.

Since $A^2>B^2$,  $A^2 \ge R^2  \ge B^2$ but  there is always  a value
$\phi_{\rm RC}$ corresponding to the case $C^2=R^2$

\begin{equation}
\tan{2\phi_{RC}}= \tan{2\delta}\,\frac{1+\cos^2{\theta}}{2\,\cos{\theta}}.
\end{equation}

\begin{figure*}[!ht]
  \centering
  \includegraphics[width=\textwidth]{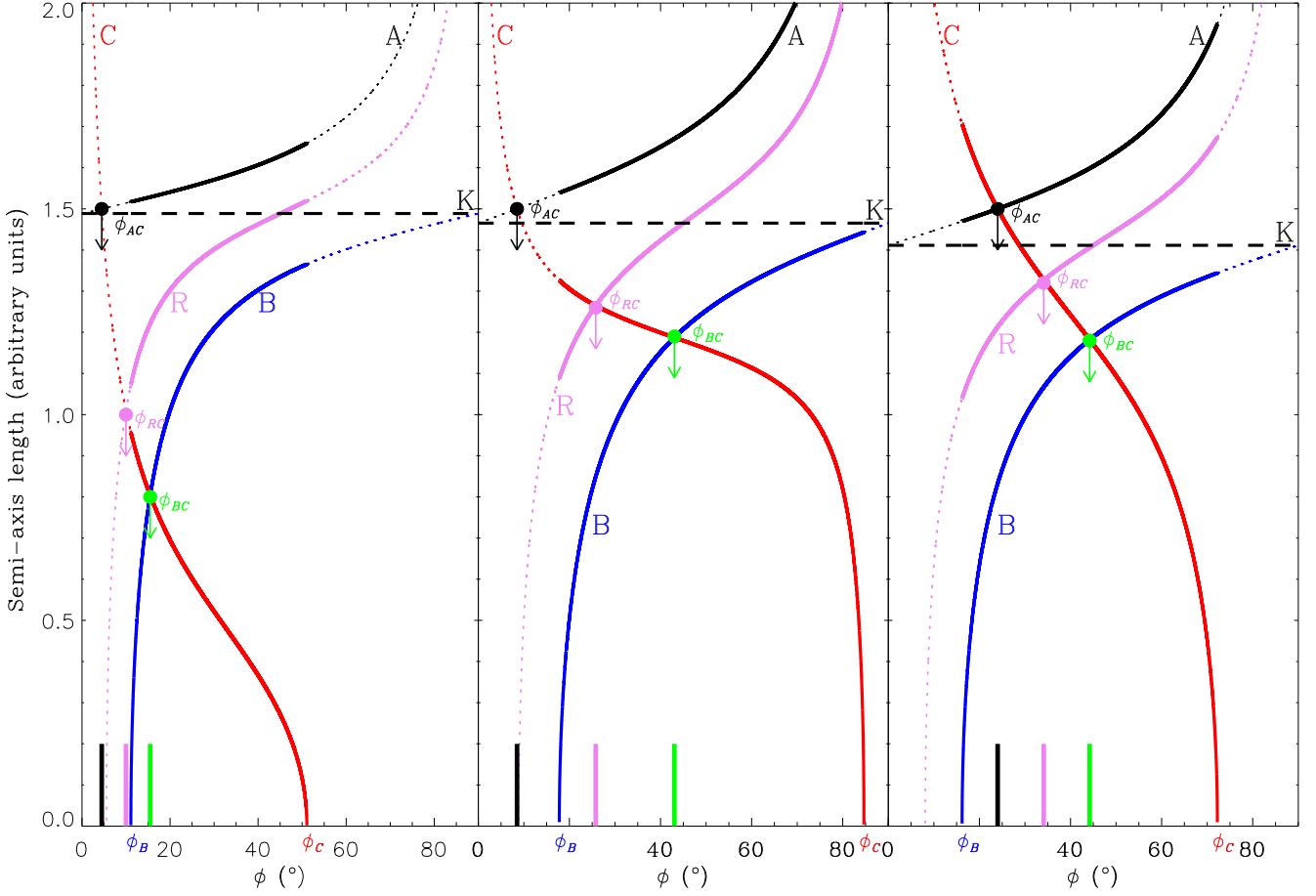}
  \caption{The lengths $A$, $B$, and $C$ of the semi-axes of the bulge
    ellipsoid and its mean equatorial  radius $R$ as a function of the
    angle  $\phi$.   The  solid  lines  correspond to  the  ranges  of
    physically possible  values of $A$,  $B$, $C$, and $R$,  while the
    dotted  lines show  their overall  trends within  $0 \le  \phi \le
    \pi/2$. A  triaxial bulge with  $\phi_{AC} < \phi_{RC}  < \phi_B$,
    $\phi_{AC}  <  \phi_B <  \phi_{RC}$,  and  $\phi_B  < \phi_{AC}  <
    \phi_{RC}$  is  shown  in  the  left, central,  and  right  panel,
    respectively.}
  \label{fig:angles}
  \end{figure*}

The mean equatorial radius allows us to distinguish oblate ($C^2<R^2$)
and  prolate  ($C^2> R^2$)  triaxial  ellipsoids.  Unfortunately,  the
situation   is  more   complicated  and   there  are   four  different
possibilities for  the intrinsic shape  of the bulge  ellipsoid.  They
are sketched in Fig.  \ref{fig:angles} and can be described as follows

\begin{itemize}

\item If  $\phi_{AC} < \phi_{RC}  < \phi_B$ the triaxial  ellipsoid is
  always  oblate (Fig.  \ref{fig:angles},  left panel).  It is  either
  completely  oblate  (i.e.,  $A>B>C$)  if $R>B>C$  ($\phi_{BC}<  \phi
  <\phi_C$)  or  partially  oblate   if  $R>C>B$  ($\phi_B  <  \phi  <
  \phi_{BC}$).

\item If $\phi_{AC}  < \phi_B < \phi_{RC}$ the  triaxial ellipsoid can
  be  either  oblate   or  prolate  (Fig.   \ref{fig:angles},  central
  panel). It is either  completely oblate if $R>B>C$ ($\phi_{BC}< \phi
  <  \phi_C$), or  partially  oblate if  $R>C>B$  ($\phi_{RC}< \phi  <
  \phi_{BC}$),  or partially  prolate  if $C>R>B$  ($\phi_B  < \phi  <
  \phi_{RC}$).

\item If $\phi_B < \phi_{AC} < \phi_{RC}$ four different possibilities
  are  allowed   for  the  triaxial  shape  of   the  bulge  ellipsoid
  (Fig. \ref{fig:angles}, right panel). It is either completely oblate
  if  $R>B>C$ ($\phi_{BC}<  \phi <  \phi_C$), or  partially  oblate if
  $R>C>B$ ($\phi_{RC}  < \phi <  \phi_{BC}$), or partially  prolate if
  $A>C>R$  ($\phi_{AC} <  \phi  < \phi_{BC}$),  or completely  prolate
  (i.e., $C>A>B$) if $C>A>R$ ($\phi_B < \phi < \phi_{AC}$).
  
\end{itemize}

\section{Equatorial ellipticity of bulges}
\label{sec:ellip}

In  Paper  I we  focused  on  the  equatorial ellipticity  defined  in
Eq. \ref{eqn:E}.  This is  a straightforward definition resulting from
the  equations  involved   in  projecting  and  deprojecting  triaxial
ellipsoids. It allows us to solve the problem of inverting an integral
equation in order  to derive the PDF of  the equatorial ellipticity of
bulges.
However, the  usual axial  ratio $B/A$ is  a more intuitive  choice to
describe the equatorial ellipticity of  the bulge when only one galaxy
is considered.   Therefore, we redefine the  equatorial ellipticity as
$Z=B^2/A^2$.   Adopting a  squared  quantity gives  us  the chance  of
successfully performing an analytic study of the problem.
By taking into account Eqs. \ref{eqn:A_K} and \ref{eqn:B_K}, we obtain

\begin{equation}
Z=\frac{B^2}{A^2}=\frac{\tan{\left(\phi-\phi_B\right)}}{\tan{\phi}}=1-\frac{2\,\sin{\phi_B}}{\sin{\phi_B} + \sin{\left(2\phi - \phi_B\right)}}.
\label{eqn:Z}
\end{equation}

$Z=0$  for  $\phi=\phi_B$,  while   the  limiting  value  of  $Z$  for
$\phi=\phi_C$ is

\begin{equation}
Z_C  =   \frac{\tan{\left(\phi_{C}-\phi_B\right)}}{\tan{\phi_{C}}}=1-\frac{2\,\sin{\phi_B}}{\sin{\phi_B} + \sin{\left(2\,\phi_C - \phi_B\right)}}.
\label{eqn:ZC}
\end{equation}
%

When $\phi$ is between $\phi_B$ and $\phi_C$, the value of $Z$ reaches
a maximum given by

\begin{equation}
Z_{\rm M} = \frac{1-\sin{\phi_B}}{1+\sin{\phi_B}} ,
\label{eqn:ZM} 
\end{equation}
%
which is observed when $\phi$ corresponds to

\begin{equation}
\phi_{\rm  M}  = \frac{\pi}{4}  +\frac{\phi_B}{2},
\end{equation}
%
where $\phi_{\rm M}$ is always larger than $\phi_B$.
The value  $Z$ decreases for  $\phi>\phi_{\rm M}$, after  reaching its
maximum   $Z_{\rm    M}$   at   $\phi=\phi_{\rm    M}$.    $Z=0$   for
$\phi=\pi/2$. But, it  is not necessary to study  the behaviour of $Z$
for  $\phi_C <  \phi \leq  \pi/2$ since  this range  of $\phi$  is not
physically possible.

Therefore, as soon as $\phi$ increases from $\phi_B$ to $\phi_C$ there
are two possible cases for  $\phi_{\rm M}$ and the corresponding trend
of $Z$. If $\phi_C>\phi_{\rm M}$, the value of $Z$ reaches the maximum
$Z_{\rm M}$ for  $\phi=\phi_{\rm M}$.  For larger values  of $\phi$ it
decreases,  reaching  the limit  value  $Z_C$  for $\phi=\phi_C$.   If
$\phi_C<\phi_{\rm M}$, $Z$  does not reach the maximum  value given by
Eq.  \ref{eqn:ZM}.  In this case, the maximum value of $Z$ corresponds
to $Z_C$.

We  also derive  for each  observed bulge  the mean  value  $\langle Z
\rangle$ of its equatorial ellipticity. From Eq. \ref{eqn:Z}

\begin{eqnarray}
\langle Z \rangle &=& \frac{1}{\phi_C-\phi_B}\int_{\phi_B}^{\phi_C} Z(\phi)\,d\phi = \nonumber \\
    &=& 1\,-\frac{\tan{\phi_B}}{\phi_C-\phi_B}\ln{\frac{\sin{\phi_C}}{\cos{\left(\phi_C-\phi_B\right)\,\sin{\phi_B}}}}.
\end{eqnarray}

  \begin{figure}[!h]
  \centering
  \includegraphics[width=0.5\textwidth]{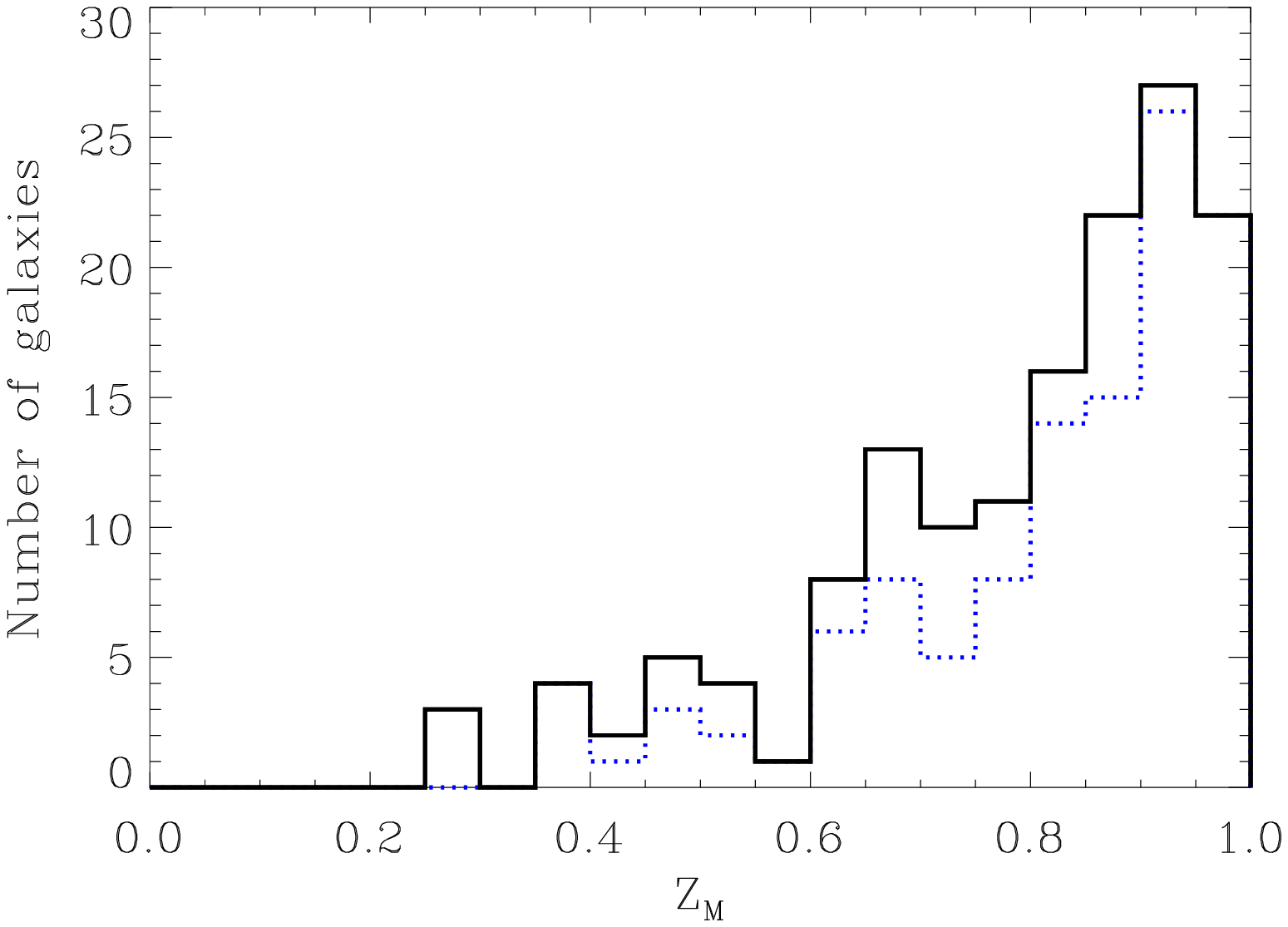}
  \includegraphics[width=0.5\textwidth]{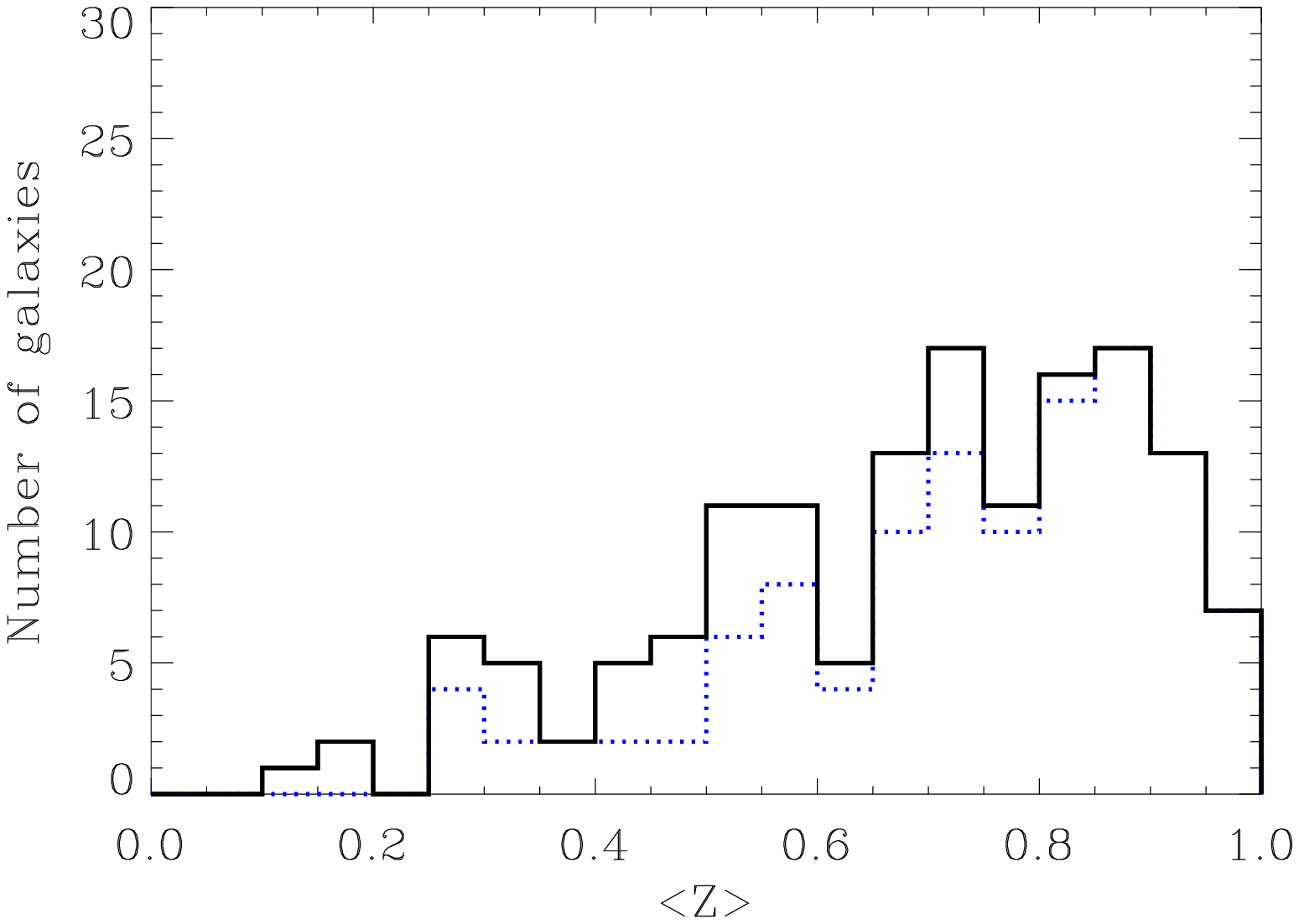}
  \includegraphics[width=0.5\textwidth]{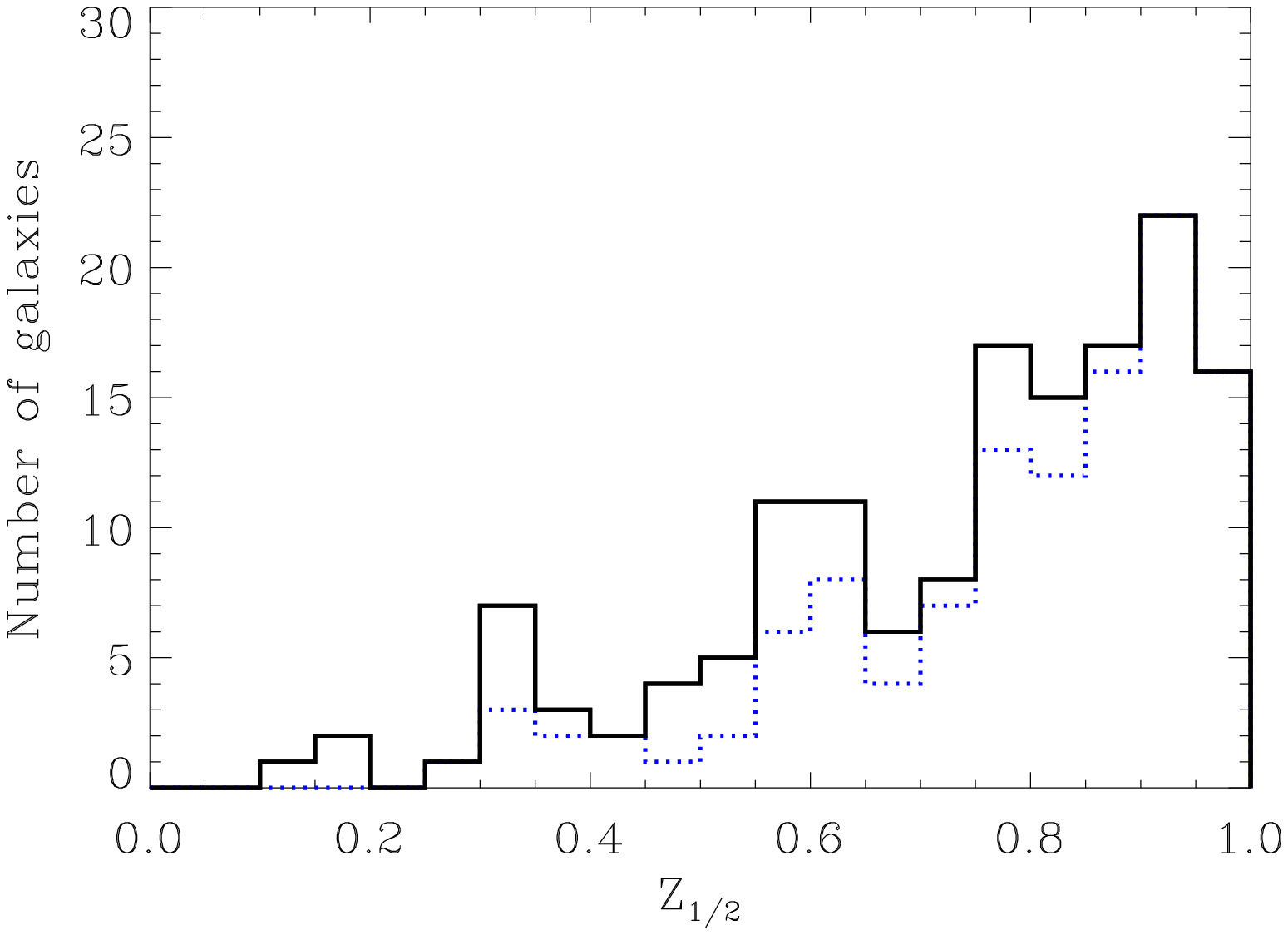}
  \caption{The distribution of the 148  sample bulges as a function of
    their maximum (top panel), mean (middle panel), and median (bottom
    panel) equatorial ellipticities plotted with a solid line. In each
    panel,  the dotted  line refers  to  the distribution  of the  115
    sample bulges with $\phi_C>\phi_{\rm M}$.}
  \label{fig:ba_statistics}%
  \end{figure}

To perform a more exhaustive statistical analysis, we compute for each
observed bulge the  probability $P(Z)$ corresponding to $0  < Z < Z_C$
by taking  into account that  $\phi$ can take  any value in  the range
$\phi_B \leq \phi \leq \phi_C$ with the same probability given by

\begin{equation}
P(\phi)=\frac{1}{\phi_C- \phi_B}.
\label{eqn:PPHI}
\end{equation}

$P(Z) = \sum P(\phi)\left|  d\phi/dZ\right|$, where the sum is defined
over all the $\phi$ values which solve Eq.  \ref{eqn:Z}.
The probability $P(Z)$ allows us to compute some characteristic values
of $Z$,  such as the median value  $Z_{1/2}$. It is defined  in such a
way  that the integrated  probability between  $Z=0$ and  $Z_{1/2}$ is
equal to the integrated probability between $Z_{1/2}$ and $Z_C$.

The distribution of the sample  bulges as a function of their maximum,
mean,    and   median   equatorial    ellipticity   is    plotted   in
Fig. \ref{fig:ba_statistics}.

Moreover, we define the confidence interval ($Z_{1/6}, Z_{5/6}$) where
the  integrated  probability  is  67\%. The  integrated  probabilities
between $Z=0$  and $Z_{1/6}$ and  between $Z=0$ and $Z_{5/6}$  are 1/6
and 5/6, respectively.
To this aim, we introduce three characteristic values of $\phi$ in the
range  between $\phi_B$  and $\phi_C$.   According to  the probability
$P(\phi)$ given in Eq. \ref{eqn:PPHI}, they are

 \begin{eqnarray}
\phi^0_{1/2}=\frac{1}{2}\phi_C + \frac{1}{2}\phi_B , \label{eqn:phi12}\\
\phi^0_{1/6}=\frac{1}{6}\phi_C + \frac{5}{6}\phi_B , \label{eqn:phi16}\\
\phi^0_{5/6}=\frac{5}{6}\phi_C + \frac{1}{6}\phi_B . \label{eqn:phi56}
\end{eqnarray}

We have seen that $Z$  has a different behaviour for $\phi_C<\phi_{\rm
  M}$ and  for $\phi_C>\phi_{\rm  M}$.  Therefore, we  will separately
study these two cases in  order to derive $P(Z)$ and the corresponding
distribution of equatorial ellipticities.

\subsection{Bulges with $\phi_C<\phi_{\rm M}$}
\label{sec:phiCsmallerphiM}

If  $\phi_C<\phi_{\rm M}$,  the value  of $Z$  monotonically increases
from $Z(\phi_{\rm  B})=0$ to $Z_C=Z(\phi_C)$. There is  only one value
of $\phi$ corresponding to any given value of $Z$.
Thus  the integrated  probability  $P(Z)$ from  $Z=0$ to  $Z=Z_{1/6}$,
$Z_{1/2}$, and $Z_{5/6}$ is equal to the integration of $P(\phi)$ from
$\phi=\phi_B$    to     $\phi=\phi^0_{1/6}$,    $\phi^0_{1/2}$,    and
$\phi^0_{5/6}$, respectively. Consequently, the median value is

\begin{equation}
Z_{1/2} = Z(\phi^0_{1/2})=1- \frac{2\,\sin{\phi_B}}{\sin{\phi_B}+\sin{\phi_C}},
\label{eqn:Z12_1}
\end{equation}
%
and the limits of the confidence interval are

\begin{equation}
Z_{1/6} = Z(\phi^0_{1/6})=1- \frac{2\,\sin{\phi_B}}{\sin{\phi_B}+\sin{\left(\frac{1}{3}\phi_C + \frac{2}{3}\phi_B\right)}},
\label{eqn:Z16_1}
\end{equation}
%
and
\begin{equation}
Z_{5/6} = Z(\phi^0_{5/6})=1- \frac{2\,\sin{\phi_B}}{\sin{\phi_B}+\sin{\left(\frac{5}{3}\phi_C - \frac{2}{3}\phi_B\right)}}.
\label{eqn:Z56_1}
\end{equation}

In this case, the probability $P(Z)$ is

\begin{equation}
P(Z) = \frac{1}{\phi_C-\phi_B} \, \frac{\sin{\phi_B}}{(1-Z)\,\sqrt{(1-Z)^2 - \sin^2{\phi_B} \,(1+Z)^2}}, \label{eqn:PZ}
\end{equation}
%
which increases monotonically between
\begin{equation}
P(0) = \frac{1}{\phi_C-\phi_B} \, \tan{\phi_B}, \label{eqn:PZ0}
\end{equation}
%
and
\begin{equation}
P(Z_C) = \frac{1}{\phi_C-\phi_B} \, \frac{1}{4} \, \frac{\left[ \sin{\phi_B} + \sin{\left(2\,\phi_C - \phi_B\right)}\right]^2}{\sin{\phi_B\,\cos{\left(2\,\phi_C - \phi_B\right)}}}.
\label{eqn:PZZC}
\end{equation}

\begin{figure}[!h]
  \centering
  \includegraphics[width=0.5\textwidth]{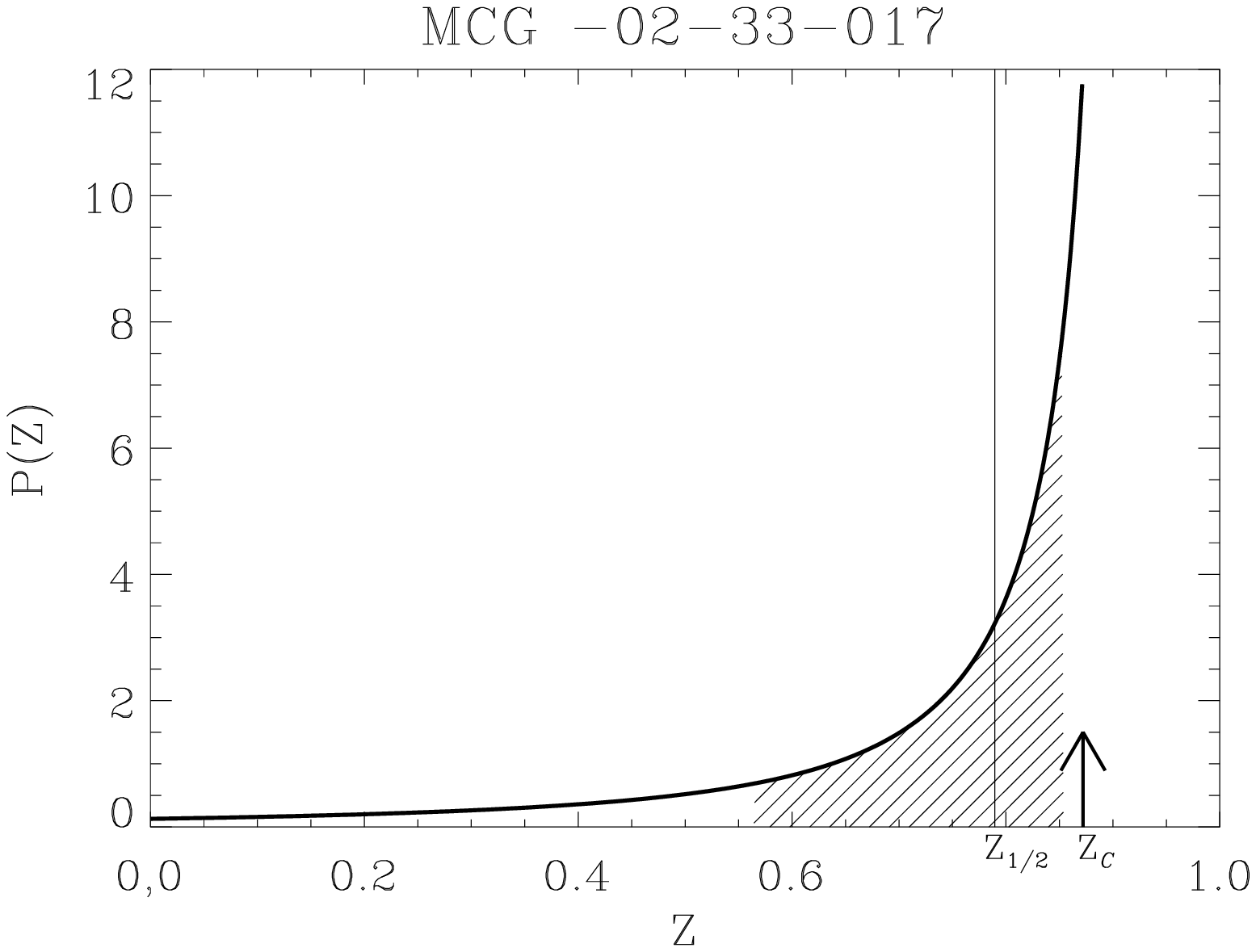}
  \includegraphics[width=0.5\textwidth]{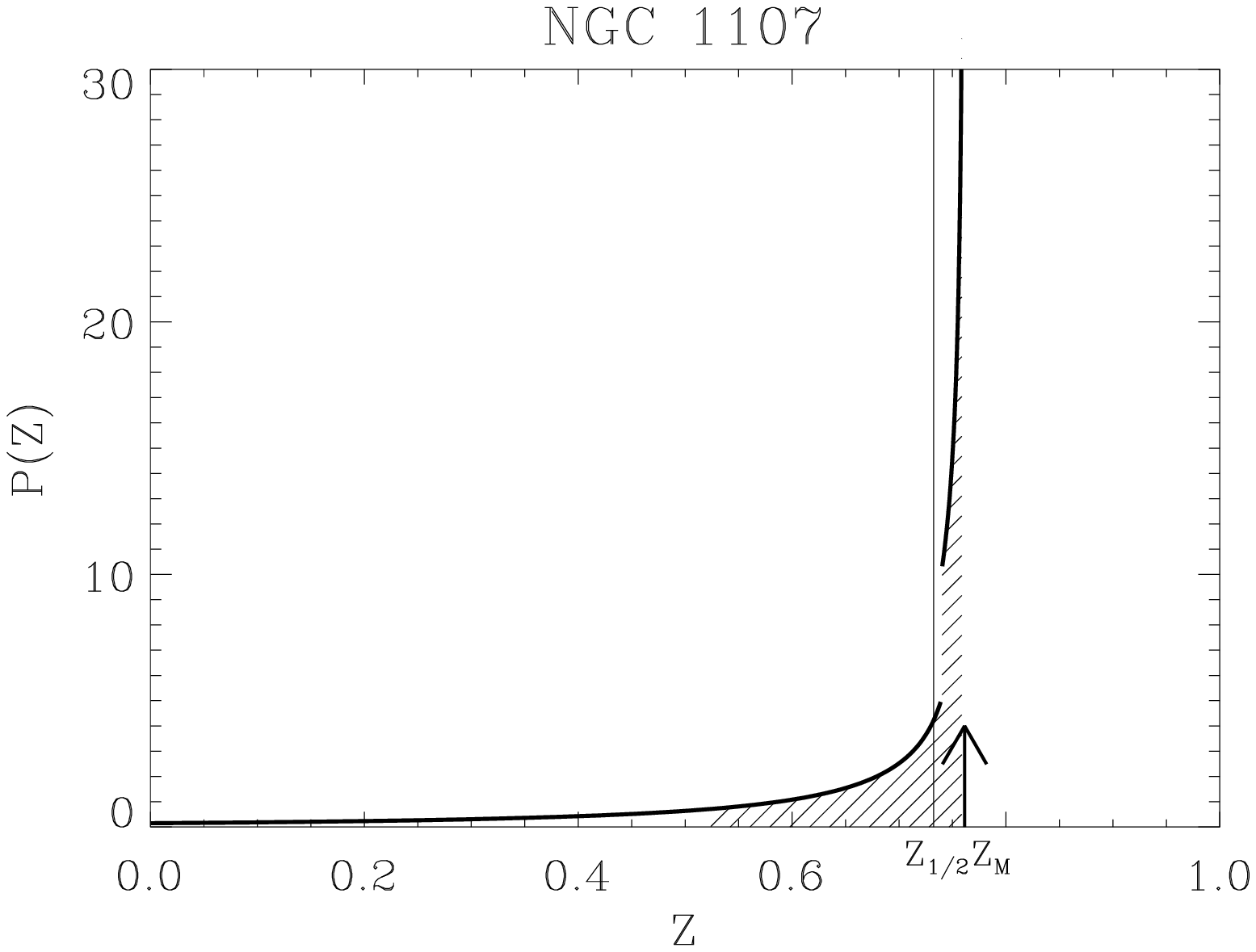}
  \includegraphics[width=0.5\textwidth]{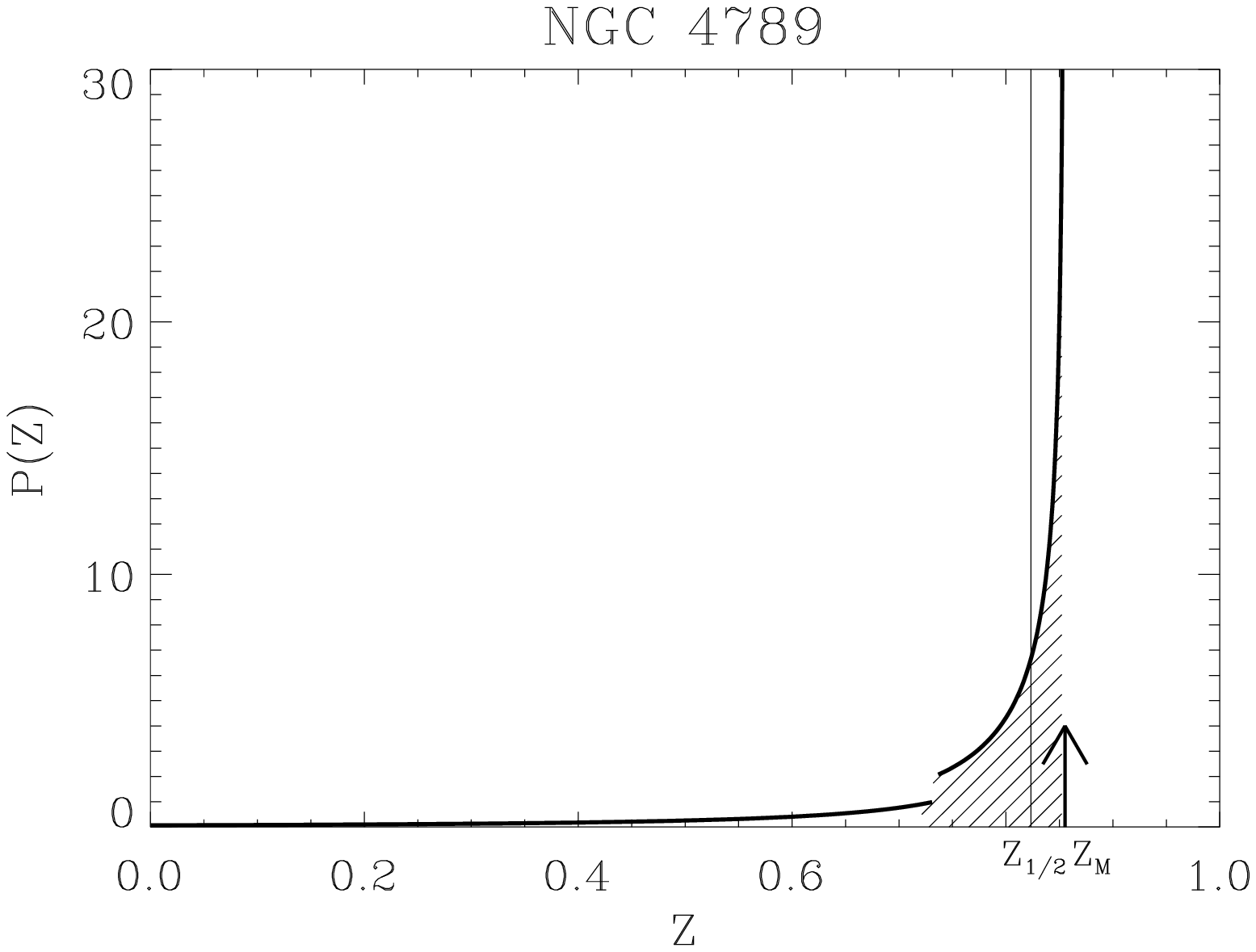}
  \caption{PDF of the equatorial  ellipticity for three sample bulges.
    MCG~-02-33-017  (top panel) hosts  a bulge  with $\phi_C<\phi_{\rm
      M}$.    NGC~1107    (middle   panel)   hosts    a   bulge   with
    $\phi_C>\phi_{\rm M}$ and $\phi^0_{1/2}<\phi'_C$. NGC~4789 (bottom
    panel)   hosts   a    bulge   with   $\phi_C>\phi_{\rm   M}$   and
    $\phi^0_{1/2}>\phi'_C$. In each panel, the vertical line shows the
    median $Z_{1/2}$ value, the arrow corresponds to the maximum value
    of  $Z$,  and  the  hatched  area marks  the  confidence  interval
    ($Z_{1/6},Z_{5/6}$) corresponding to $67\%$ probability.}
  \label{fig:prob_ba}
  \end{figure}

The probability  $P(Z)$ given in  Eq.  \ref{eqn:PZ} strongly  peaks at
$Z=Z_C$ in  such a  way that  $Z_{1/2}$ is close  to $Z_C$.   For this
reason,  although   the  right  portion   ($Z_{1/2},Z_{5/6}$)  of  the
confidence interval  ($Z_{1/6},Z_{5/6}$) is not  large, the confidence
interval  spans a  large fraction  of the  total range  between  0 and
$Z_C$.   This  is the  case  for  the  bulge of  MCG~-02-33-017  (Fig.
\ref{fig:prob_ba}, top panel). Using  the mean $\langle Z \rangle$ and
median  $Z_{1/2}$ values  to  describe the  equatorial ellipticity  of
these kinds of bulges is a poor approximation.

\subsection{Bulges with $\phi_C>\phi_{\rm M}$}
\label{sec:phiClargerphiM}

For   $\phi_C>\phi_{\rm   M}$,   $Z$  monotonically   increases   from
$Z(\phi_B)=0$ to $Z_{\rm M}=Z(\phi_{\rm M})$ and then it monotonically
decreases from $Z_{\rm M}$ to $Z_C=Z(\phi_C)$.  For $0<Z<Z_C$ there is
only  one  value   of  $\phi$  for  each  value   of  $Z$,  while  for
$Z_C<Z<Z_{\rm M}$ there  are two values of $\phi$  which correspond to
each value  of $Z$.  There is a  discontinuity in $P(Z)$  for $Z=Z_C$,
which   corresponds   to    the   value   $\phi'_{C}=\frac{\pi}{2}   -
(\phi_C-\phi_B)$.  $dZ/d\phi  = 0$  for  $\phi=\phi_{\rm  M}$ and  the
probability $P(Z)$ becomes infinity at $Z=Z_{\rm M}$.
It is not possible to  compute directly the median value $Z_{1/2}$ and
confidence    interval   ($Z_{1/6},Z_{5/6}$)    from    $P(\phi)$   in
Eq. \ref{eqn:PPHI}. Therefore, we need to rewrite $P(Z)$ as

\begin{eqnarray}
P(Z)  = \left \{ \begin{array}{l} \frac{1}{\phi_C-\phi_B} \, \frac{\sin{\phi_B}}{(1-z)\,\sqrt{(1-z)^2 - \sin^2{\phi_B} \,(1+z)^2}} \qquad 0 \le Z < Z_C, \label{eqn:PZ1}\\
     \frac{2}{\phi_C-\phi_B} \, \frac{\sin{\phi_B}}{(1-z)\,\sqrt{(1-z)^2 - \sin^2{\phi_B} \,(1+z)^2}} \qquad Z_C \le Z \le Z_{\rm M}. \label{eqn:PZ2}\\
     \end{array} \right.
\end{eqnarray}

There  are different  values for  $Z_{1/6}$, $Z_{1/2}$,  and $Z_{5/6}$
depending  on  whether  $\phi^0_{1/2}$  is  smaller  or  greater  than
$\phi'_C$ which corresponds to the discontinuity in $P(Z)$.

For $\phi^0_{1/2} < \phi'_C$ the values of $Z_{1/2}$ and $Z_{1/6}$ are
given  by  Eqs.   \ref{eqn:Z12_1} and  \ref{eqn:Z16_1},  respectively.
But,  there are  two possible  values for  $Z_{5/6}$ depending  on the
value of $\phi^0_{5/6}$.   If $\phi^0_{5/6}<\phi'_C$ then $Z_{5/6}$ is
given  by  the Eq.   \ref{eqn:Z56_1}.   If $\phi^0_{5/6}>\phi'_C$  the
corresponding values of $Z$ are on the right side of the discontinuity
(i.e.,  two values  of $Z$  correspond  to a  given value  of $\phi  >
\phi'_C$). In this case

\begin{equation}
Z_{5/6} = 1- \frac{2\,\sin{\phi_B}}{\sin{\phi_B}+\cos{\frac{\phi_C - \phi_B}{6}}},
\label{eqn:Z56_2}
\end{equation}
%
which corresponds to $Z(\phi_{5/6})$ with $\phi_{5/6}=\pi/4-\phi_C/12
+ 7\,\phi_B/12$.

For $\phi^0_{1/2} > \phi'_C$ the value of $Z_{1/2}$ is given by

\begin{equation}
Z_{1/2} = 1- \frac{2\,\sin{\phi_B}}{\sin{\phi_B}+\cos{\frac{\phi_C -
      \phi_B}{2}}},
\label{eqn:Z12_2}
\end{equation}
%
and it corresponds  to $Z(\phi_{1/2})$ with $\phi_{1/2}=\pi/2-\phi_C/2
+ 3\,\phi_B/2$.  Likewise, $Z_{5/6}$ is given by Eq.  \ref{eqn:Z56_2}.
But, for $Z_{1/6}$ we have two possibilities according to the value of
$\phi^0_{1/6}$.  If $\phi^0_{1/6}<\phi'_C$  then $Z_{1/6}$ is given by
Eq.   \ref{eqn:Z16_1}.   If  $\phi^0_{1/6}>\phi'_C$ the  corresponding
values of $Z$ are on the right side of the discontinuity, and it is

\begin{equation}
Z_{1/6} = 1- \frac{2\,\sin{\phi_B}}{\sin{\phi_B}+\cos{\frac{5\,\left(\phi_C - \phi_B\right)}{6}}}
\label{eqn:Z16_2}
\end{equation}
%
which corresponds to $Z(\phi_{1/6})$ with
$\phi_{1/6}=\pi/4-5\,\phi_C/12 +11\,\phi_B/12$.

For $\phi_C>\phi_{\rm M}$ the probability $P(Z)$ in Eq.  \ref{eqn:PZ1}
peaks strongly at  $Z_{\rm M}$ and therefore the  median $Z_{1/2}$ and
maximum  $Z_{\rm M}$  values of  the equatorial  ellipticity  are very
close and the confidence interval ($Z_{1/6},Z_{5/6}$) is narrow.  This
is  the case  for  the bulges  of  NGC~1107 (Fig.   \ref{fig:prob_ba},
middle  panel) and NGC~4789  (Fig.  \ref{fig:prob_ba},  bottom panel).
We  conclude that for  these types  of bulges  the statistics  we have
presented  here  are  representative  of  their  intrinsic  equatorial
ellipticity.

\subsection{Statistics of the equatorial ellipticity of bulges}

The distribution of  the maximum equatorial ellipticity (corresponding
to either $Z_C$  for bulges with $\phi_C<\phi_{\rm M}$  or $Z_{\rm M}$
for  bulges  with  $\phi_C>\phi_{\rm  M}$) peaks  at  $Z_{\rm  M}>0.9$
(Fig. \ref{fig:ba_statistics},  top panel). These  are nearly circular
bulges ($B/A=0.95$).   But, we conclude  that a large fraction  of the
sample bulges are  strong candidates to be triaxial  because $41\%$ of
them have $Z_{\rm M} < 0.80$ ($B/A<0.89$).
This result is  in agreement with our previous finding  in Paper I and
with   the    analysis   of    the   distribution   of    mean   (Fig.
\ref{fig:ba_statistics},      middle       panel)      and      median
(Fig. \ref{fig:ba_statistics}, bottom  panel) ellipticities.  In fact,
we find that $64\%$ and $53\%$ of our bulges have $\langle Z \rangle <
0.8$ and $Z_{1/2} < 0.8$,  respectively. The mean values of $\langle Z
\rangle$ and $Z_{1/2}$ are 0.68 and 0.73, respectively.

The width of the confidence interval ($Z_{1/6},Z_{5/6}$) corresponding
to  a  $67\%$  probability is  related  to  the  accuracy of  the  $Z$
measurement. The  narrowest confidence intervals are  found for bulges
with  $\phi_C \rightarrow  \pi/2$  and $\phi_B  \rightarrow 0$.   This
implies that  $\phi'_C \rightarrow  \phi_B$ and $Z_C>Z_{\rm  M}$.  For
these bulges  the discontinuity in  $P(Z)$ is almost  negligible.  The
case  with   $\phi_C=\pi/2$  and  $\phi_B=0$   corresponds  either  to
spherical bulges (i.e., $e=0$) or to bulges with a circular equatorial
section (i.e.,  $\tan{2\delta}=0$).  Consequently, the  bulges with $B
\approx A$  are among those  characterized by the  narrower confidence
interval and better determination of $Z$.
We can select all sample objects for which the $Z$ measurement is only
slightly uncertain.  They are  the 115 galaxies with $\phi_C>\phi_{\rm
  M}$.  The distribution  of these  selected bulges  as a  function of
their $Z_{\rm  M}$, $\langle Z  \rangle$, and $Z_{1/2}$ is  plotted in
Fig.  \ref{fig:ba_statistics} too.
The  fraction  of  bulges  with  $Z_{\rm M}<0.8$  is  $33\%$.   It  is
significantly smaller  than the $41\%$ found for  the complete sample,
because the selected sample is biased toward bulges with $B \approx A$
including  all  the  bulges  with  a  circular  (or  nearly  circular)
equatorial section.
The fraction of selected bulges with $\langle Z \rangle < 0.8$ and
$Z_{1/2} < 0.8$ is $55\%$ and $43\%$, respectively.

\section{Intrinsic flattening of bulges}
\label{sec:flat}

The axial  ratio $C/A$ usually describes the  intrinsic flattening $F$
of  a triaxial  ellipsoid if  $A \geq  B \geq  C$.  Since  we  have no
constraints  on  the  lengths  $A$,  $B$, and  $C$,  we  redefine  the
flattening as

\begin{equation}
F(\phi)=\frac{C^2}{R^2}=\frac{2 C^2}{A^2+B^2},
\label{eqn:Fdef}
\end{equation}
%
by using the  lengths $C$ and $R$ of the polar  semi-axis and the mean
equatorial  radius   given  by  Eqs.    \ref{eqn:C}  and  \ref{eqn:R},
respectively.

\begin{equation}
F(\phi)=F_{\theta}\,\frac{\cot{2\phi}-\cot{2\phi_C}}{\cot{\phi_B} - \cot{2\phi}}=F_{\theta}\,\frac{\sin{\phi_B}}{\sin{2\phi_C}} \frac{\sin{\left(2\phi_C - 2\phi\right)}}{\sin{\left(2\phi - \phi_B\right)}},
\label{eqn:F}
\end{equation}
%
where
 
\begin{equation}
F_{\theta}=\frac{2\,\cos^2{\theta}}{\sin^2{\theta}},
\label{eqn:Ftheta}
\end{equation}
%
accounts for the effect of inclination. The angle $\theta$ also enters
in  the definition of  the two  angles $\phi_B$  and $\phi_C$  in Eqs.
\ref{eqn:B=0} and \ref{eqn:C=0}, respectively.
Adopting  a  squared   quantity  for  $F$  gives  us   the  chance  of
successfully performing an  analytic study of the problem  as was done
for the equatorial ellipticity $Z$ in Eq. \ref{eqn:Z}.

Since  $dF(\phi)/d\phi<0$,  the  function $F(\phi)$  is  monotonically
decreasing with a maximum $F_{\rm M}$ at $\phi=\phi_B$ given by

\begin{equation}
F_{\rm M}=F_{\theta}\frac{\sin{\left(2\,\phi_C -2\,\phi_B\right)}}{\sin{2\,\phi_C}}.
\label{eqn:FM}
\end{equation}

If $\phi$ increases from $\phi_B$ to $\phi_C$, the value of $F(\phi)$
decreases to zero at $\phi=\phi_C$.
According  to  Eq.   \ref{eqn:FM},  for  $F_{\rm  M}<1$  the  triaxial
ellipsoids are  oblate, with some  of them being partially  oblate and
others completely  oblate. For  $F_{\rm M}>1$ the  triaxial ellipsoids
can also  be partially  prolate and in  some extreme  cases completely
prolate.

From Eq. \ref{eqn:F}, we compute the mean value $\langle F \rangle$ of
the intrinsic flattening as

\begin{eqnarray}
\langle F \rangle &=& \frac{1}{\phi_C-\phi_B}\int_{\phi_B}^{\phi_C} F(\phi)\,d\phi = \nonumber \\
                  &=& F_{\theta}\frac{\sin{\phi_B}}{\sin{2\,\phi_C}} \left[ \frac{\sin{\left(2\,\phi_C-\phi_B\right)}}{2\,\left(\phi_C - \phi_B\right)} \ln{\left(\frac{\sin{\left(2\,\phi_C-\phi_B\right)}}{\sin{\phi_B}}\right)} - \right. \nonumber \\
                  & & \left.  - \cos{\left(2\,\phi_C - \phi_B\right)}\right].
\end{eqnarray}

Since  $F(\phi)$ is  a monotonic  function  (i.e., each  value of  $F$
corresponds to  only one value of $\phi$),  the integrated probability
$P(F)$   between   $F(\phi_C)=0$   and   some   characteristic   value
$F_{*}=F(\phi_{*})$  is equal  to  the integral  of $P(\phi)$  between
$\phi_{*}$ and  $\phi_C$.  Then, it is straightforward  to compute the
median value  $F_{1/2}$ of the intrinsic  flattening which corresponds
to the median value $\phi^0_{1/2}=(\phi_C+\phi_B)/2$

\begin{equation}
F_{1/2}=F_{\theta}\,\frac{\sin{\phi_B}}{\sin{\left(2\,\phi_C\right)}} \frac{\sin{\left(\phi_C-\phi_B\right)}}{\sin{\phi_C}}.
\end{equation}

\begin{figure}[!h]
  \centering
  \includegraphics[width=0.5\textwidth]{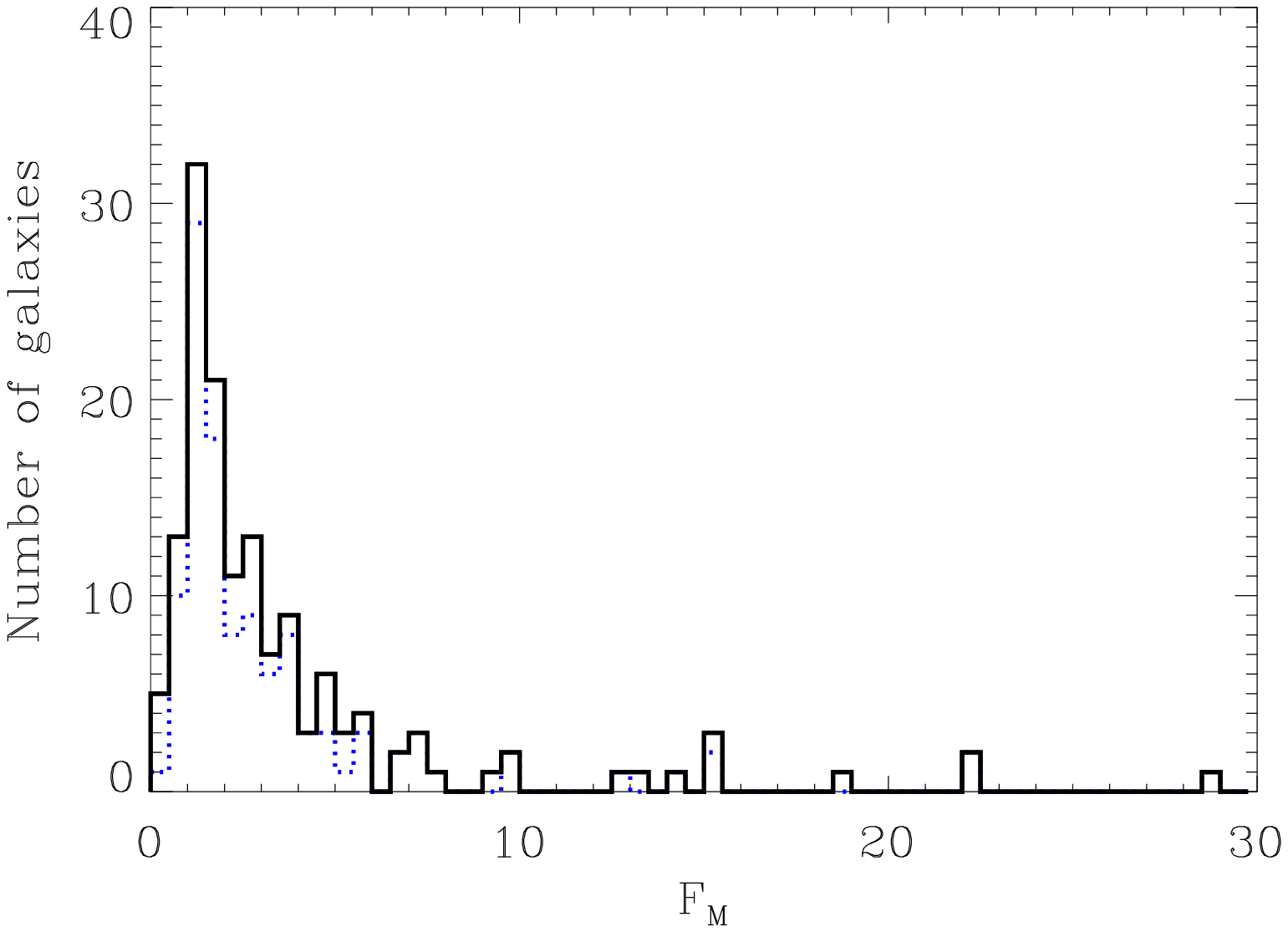}
  \includegraphics[width=0.5\textwidth]{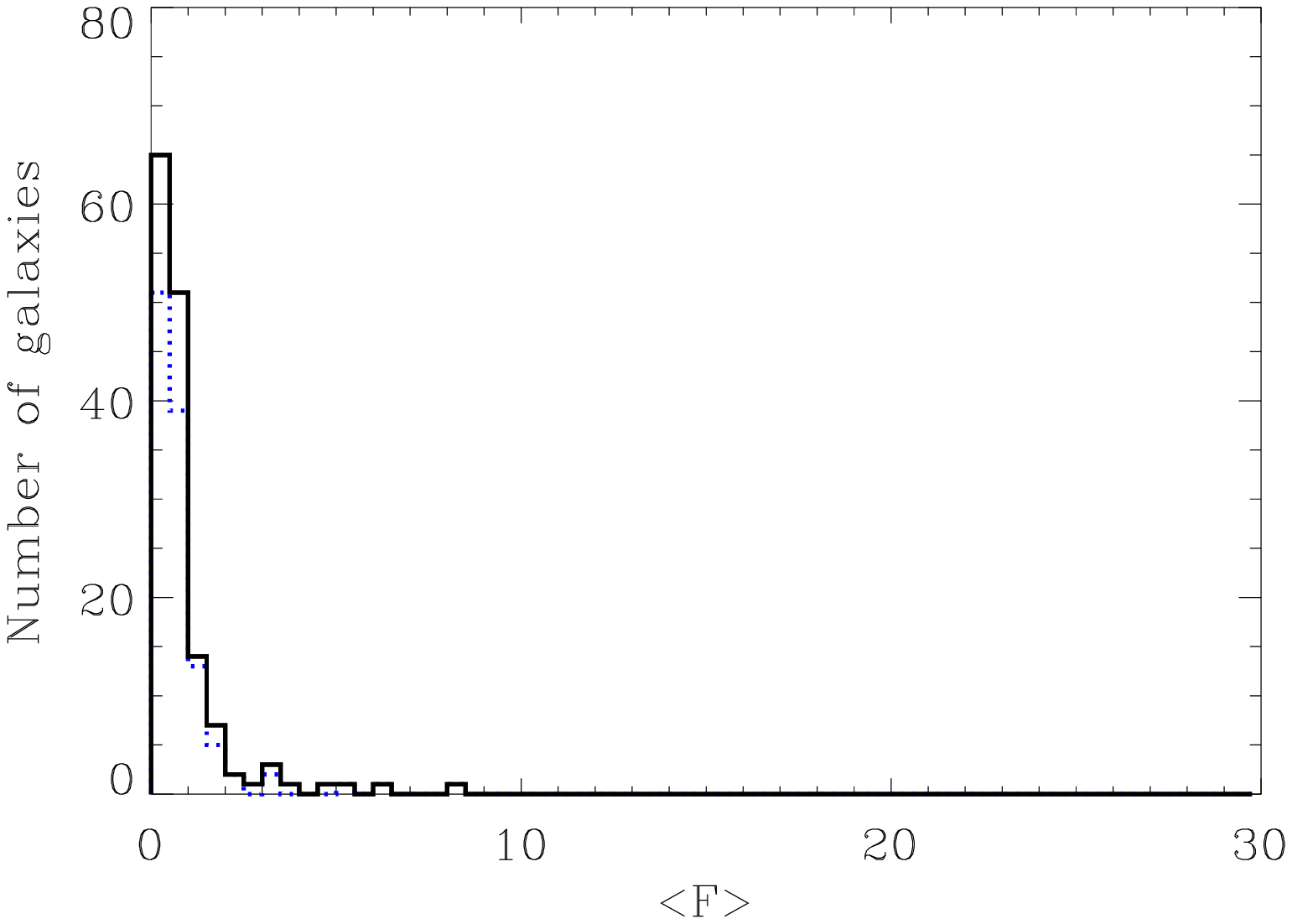}
  \includegraphics[width=0.5\textwidth]{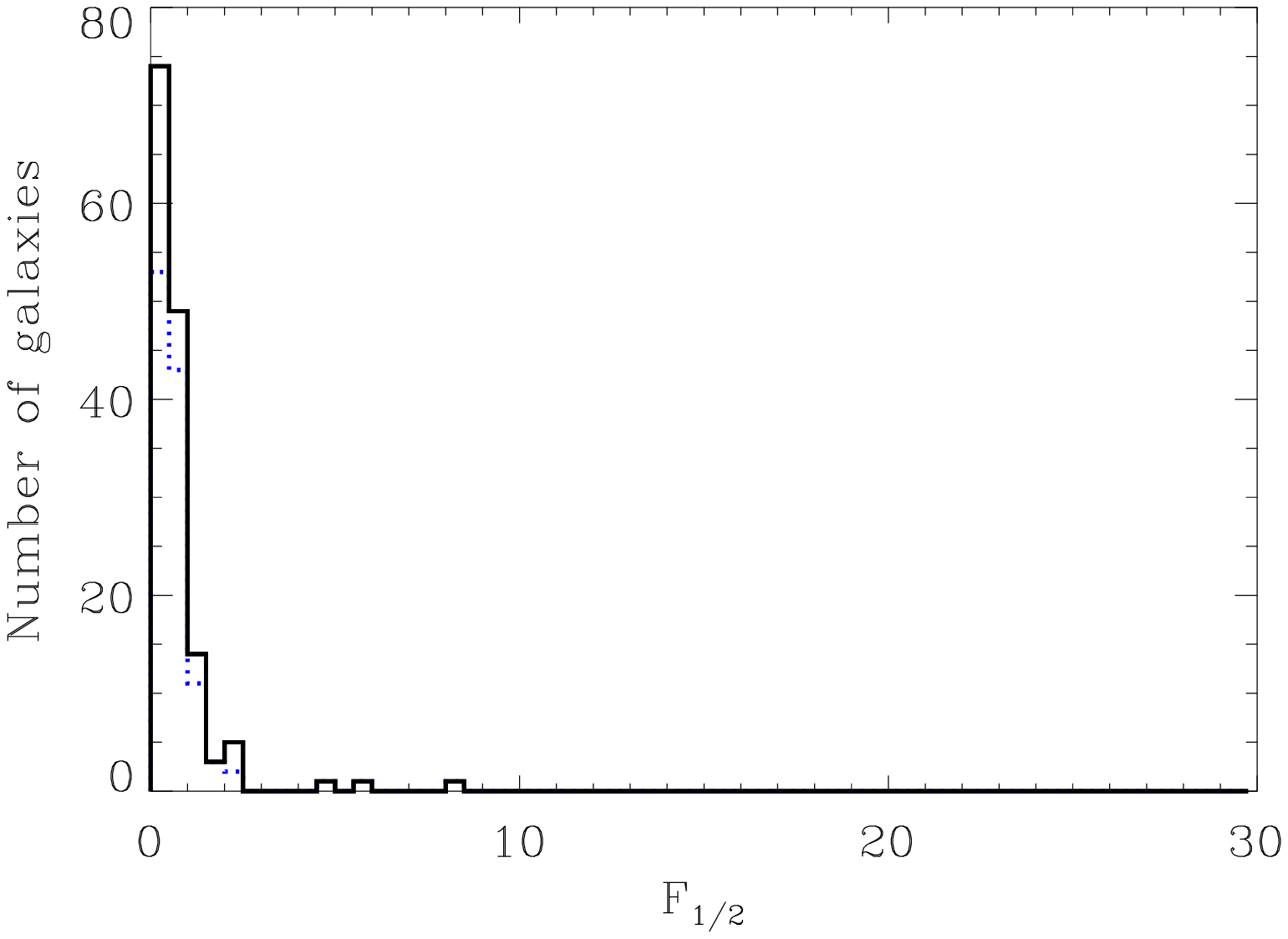}
  \caption{The distribution of the 148  sample bulges as a function of
    their maximum (top panel), mean (middle panel), and median (bottom
    panel) intrinsic  flattening, plotted with  a solid line.  In each
    panel,  the dotted  line refers  to  the distribution  of the  115
    sample bulges with $\phi_C>\phi_{\rm M}$.}
  \label{fig:flatstat}%
  \end{figure}

As was done for the equatorial ellipticity, we can define also for the
flattening  a  confidence interval  ($F_{1/6}$,  $F_{5/6}$) where  the
integrated   probability   is  $67\%$.    In   fact,  the   integrated
probabilities  between  $F=0$  and  $F_{1/6}$ and  between  $F=0$  and
$F_{5/6}$ are 1/6 and 5/6, respectively.
We have 

\begin{equation}
F_{1/6}=F_{\theta}\, \frac{\sin{\phi_B}}{\sin{\left(2\,\phi_C\right)}} \frac{\sin{\left(\frac{1}{3}\phi_C-\frac{1}{3}\phi_B\right)}}{\sin{\left(\frac{5}{3}\phi_C- \frac{2}{3}\phi_B\right)}},
\end{equation}
%
which corresponds to $\phi^0_{1/6}$ given in Eq. \ref{eqn:phi16}, and

\begin{equation}
F_{5/6}=F_{\theta}\,\frac{\sin{\phi_B}}{\sin{\left(2\,\phi_C\right)}} \frac{\sin{\left(\frac{5}{3}\phi_C-\frac{5}{3}\phi_B\right)}}{\sin{\left(\frac{1}{3}\phi_C+ \frac{2}{3}\phi_B\right)}},
\end{equation}
%
which corresponds to $\phi^0_{5/6}$ given in Eq. \ref{eqn:phi56}.
The distribution of the sample  bulges as a function of their maximum,
mean,    and    median   intrinsic    flattening    is   plotted    in
Fig. \ref{fig:flatstat}.

It is possible to perform a more exhaustive statistical analysis by
defining the probability $P(F)$ of having a flattening $F$ as

\begin{equation}
P(F)=k_{0}\,\frac{1}{A_{0}F^2 + B_{0}F + C_{0}},
\end{equation}
%
where
\begin{eqnarray}
k_{0} &=& \frac{\cos^2{\theta}\,\sin{\left(2\phi_C-\phi_B\right)}}{\sin^2{\theta}\,(\phi_C-\phi_B)\,\sin{2\phi_C}\, \sin{\phi_B}},\\
A_{0} &=& \frac{1}{\sin^2{\phi_B}},\\
B_{0} &=& \frac{4\,\cos^2{\theta}\,\cos{\left(2\phi_C-\phi_B\right)}}{\sin^2{\theta}\,\sin{2\phi_C}\,\sin{\phi_B}},\\
C_{0} &=& \frac{4\,cos^4{\theta}}{\sin^4{\theta}\,\sin^2{2\phi_C}},\\
\end{eqnarray}
%
where $k_{0}$, $A_{0}$, and  $C_{0}$ are always positive, while $B_{0}
> 0$ for $2\phi_C-\phi_B<\pi/2$ ($\phi_C < \phi_{\rm M}$) and $B_{0} <
0$ for  $2\phi_C-\phi_B>\pi/2$ ($\phi_C  > \phi_{\rm M}$).   All these
quantities can  be computed directly  for each observed  bulge, indeed
they depend  only on  the measured values  of $a$, $b$,  $\delta$, and
$\theta$ through the angles $\phi_B$ and $\phi_C$.

\begin{figure*}[!ht]
  \centering
  \includegraphics[width=0.48\textwidth]{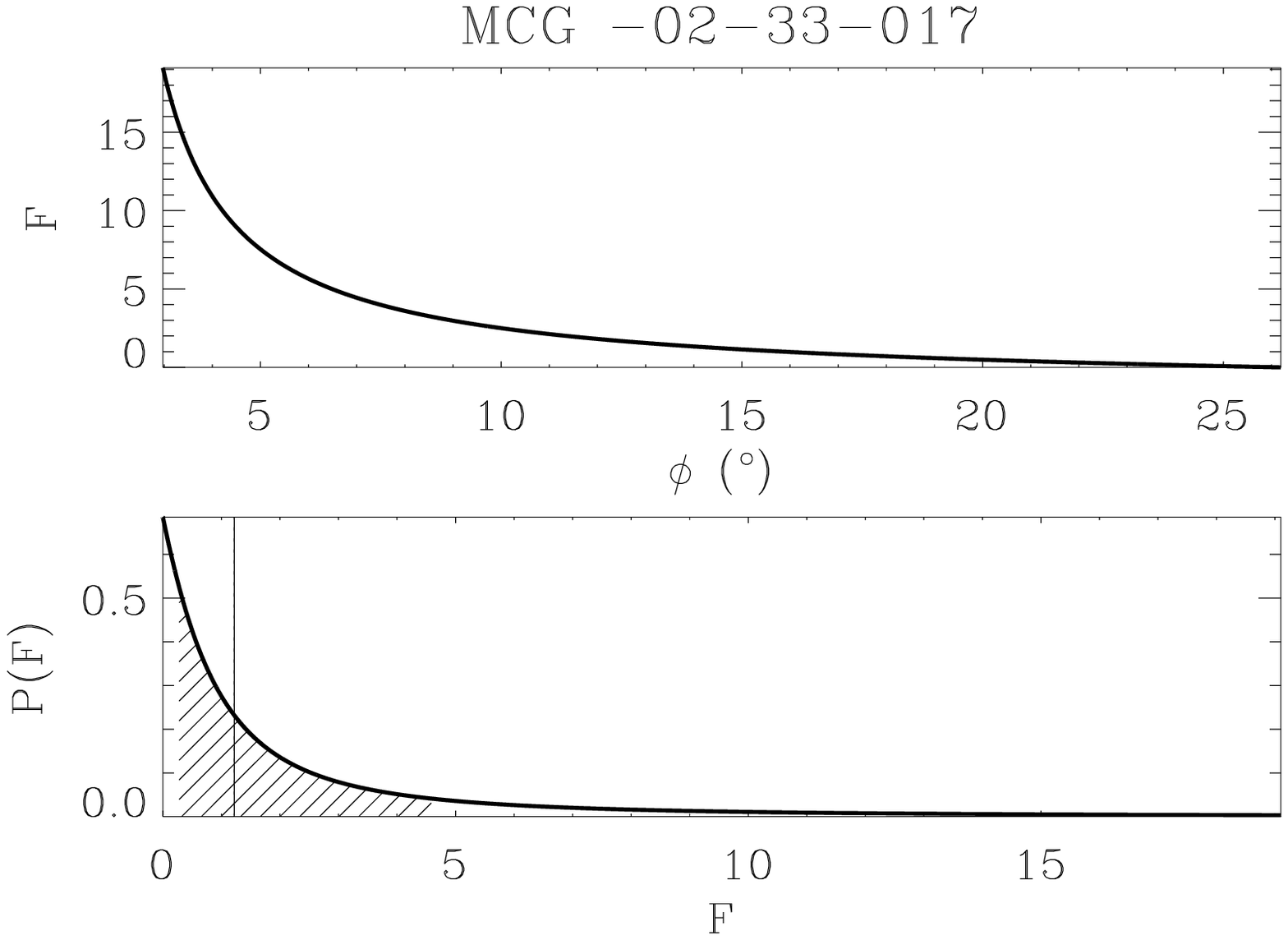}
  \includegraphics[width=0.48\textwidth]{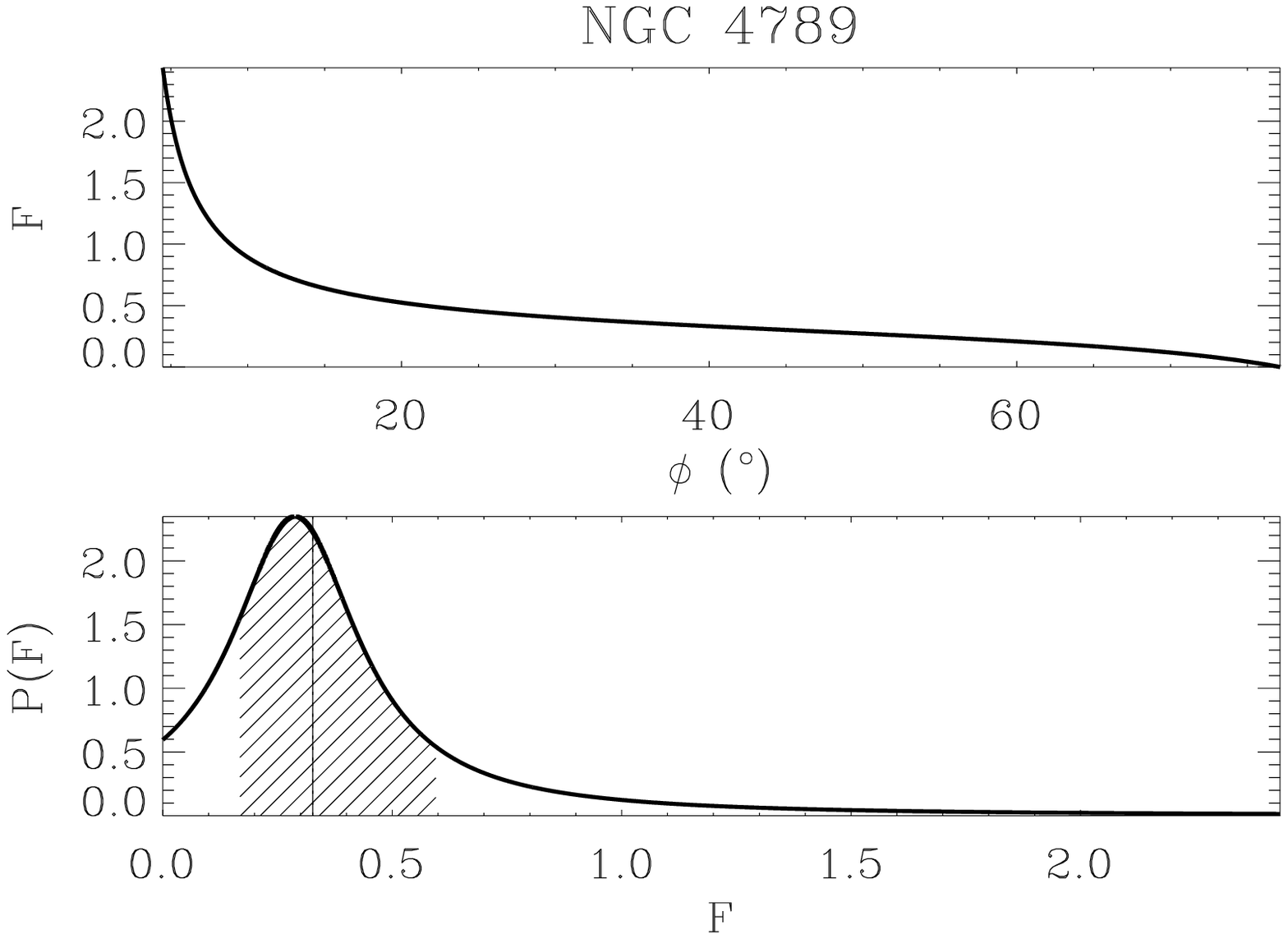}
  \caption{The intrinsic flattening as  a function of the angle $\phi$
    (top panels)  and its PDF  (bottom panels) for two  sample bulges.
    MCG~-02-33-017 (left  panels) and  NGC~4789 (right panels)  host a
    bulge with  $B_{0}>0$ and $B_{0}<0$, respectively.   In the bottom
    panels, the vertical line shows the median $F_{1/2}$ value and the
    hatched  area marks  the  confidence interval  ($F_{1/6},F_{5/6}$)
    corresponding to $67\%$ of probability.}
  \label{fig:flatprob}%
  \end{figure*}

In  Sect.   \ref{sec:phiCsmallerphiM}  we  found that  the  confidence
interval ($Z_{1/6}$, $Z_{5/6}$) of  equatorial ellipticity for a bulge
with  $\phi_C < \phi_{\rm  M}$ is  wide. For  this reason,  the median
$Z_{1/2}$ and  mean $\langle Z \rangle$ values  are not representative
of  the equatorial ellipticity  of the  bulge.  The  same is  true for
($F_{1/6}$, $F_{5/6}$)  because the probability  function $P(F)$ peaks
at  $F=0$  and  slowly decreases  as  soon  as  $F$ increases.   As  a
consequence, the median $F_{1/2}$  and mean $\langle F \rangle$ values
are not representative of the intrinsic flattening of the bulge.
This    is    the   case    for    the    bulge   of    MCG~-02-33-017
(Fig. \ref{fig:flatprob}, left panels)

On  the contrary,  if $\phi_C>  \phi_{\rm  M}$ then  $B_0<0$, and  the
probability function $P(F)$ peaks at the most probable value

\begin{equation}
F_{\rm MP}=-\frac{1}{2}\frac{B_{0}}{A_{0}},
\end{equation}
%
and it quickly decreases to

\begin{equation}
P(0)=\frac{k_{0}}{C_{0}},
\end{equation}
%
and to zero for $F<F_{\rm  MP}$ and $F>F_{\rm MP}$, respectively.  The
confidence  interval  ($F_{1/6}$, $F_{5/6}$)  is  narrow.  The  median
$F_{1/2}$,  mean $\langle  F  \rangle$, and  the  most probable  value
$F_{\rm  MP}$   are  close  to  each   other  and  all   of  them  are
representative of the intrinsic flattening.
This is the  case for the bulge of  NGC~4789 (Fig. \ref{fig:flatprob},
right panels).

\subsection{Statistics of the intrinsic flattening of bulges}

\begin{figure}[!h]
  \centering
  \includegraphics[width=0.5\textwidth]{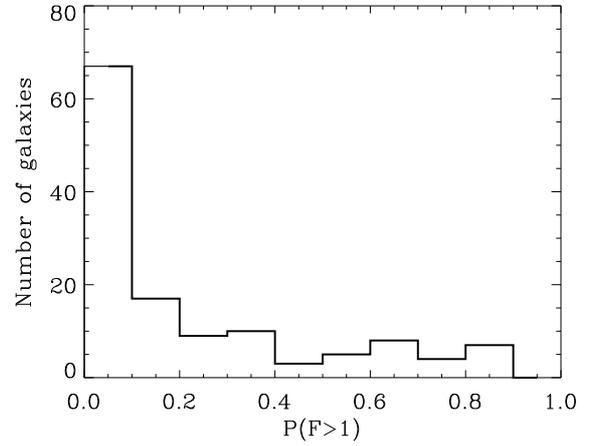}
  \caption{Number  of  sample bulges  which  could  have an  intrinsic
    flattening  $F>1$  as a  function  of  the  probability that  this
    happens. Bulges with $F<1$ (i.e., oblate triaxial ellipsoids) have
    been not taken into account.}
  \label{fig:pfgt1}%
  \end{figure}

The    distribution    of    the    maximum    intrinsic    flattening
(Fig. \ref{fig:flatstat},  top panel) shows that $12\%$  of the sample
bulges  have  $F_{\rm  M}<1$  (i.e.,  they are  either  completely  or
partially oblate triaxial ellipsoids).
Judging by $F_{\rm M}$, the  majority of sample bulges could be highly
elongated along the polar axis. However, these highly elongated bulges
are not common.  Indeed, after  excluding from the complete sample the
bulges with $F_{\rm M}<1$, only $19\%$ ($18\%$ if we consider only the
selected  sample  of  115  bulges)  of the  remaining  bulges  have  a
probability greater than $50\%$  to have an intrinsic flattening $F>1$
and there are no bulges with  more than a $90\%$ probability of having
$F>1$ (Fig.  \ref{fig:pfgt1}).
This is  in agreement with  the results based  on the analysis  of the
distribution of  the mean (Fig. \ref{fig:flatstat},  middle panel) and
median (Fig.  \ref{fig:flatstat}, bottom panel) intrinsic flattening.
We find that $78\%$ of the sample bulges have $\langle F \rangle < 1$,
and $83\%$ have $F_{1/2} < 1$. They are oblate triaxial ellipsoids.

The large number of sample bulges  with $F_{\rm M} >1$ with respect to
those which  are actually elongated along  the polar axis is  due to a
projection effect of the triaxial ellipsoids.
For any $\phi$  the contribution of inclination $\theta$  to the value
of    $F$    is    given     by    $F_{\theta}$    as    defined    in
Eq.  \ref{eqn:Ftheta}. However, the  intrinsic flattening  scales with
$F_{\theta}$,   whereas    the   probability   $P(F)$    scales   with
$1/F_{\theta}$.   Thus,  the probability  to  have large  $F_{\theta}$
values (and large $F_{\rm M}$ values) is very small. For instance, the
probability  to   have  the  maximum   $F_{\rm  M}$  value   given  by
Eq. \ref{eqn:FM} is

\begin{equation}
P(F_{\rm M})=\frac{1}{2\,\left(\phi_C - \phi_B\right)}\frac{1}{F_{\theta}}\frac{\sin{\phi_B}\,\sin{2\,\phi_C}}{\sin{\left(2\,\phi_C - \phi_B\right)}}.
\end{equation}
%
We conclude  that  $F_{\rm M}$  is  not a  good  proxy for  the
intrinsic flattening of  a bulge, although $Z_{\rm  M}$ is a good
  proxy for equatorial ellipticity.

The distribution of the  selected bulges with $\phi_C>\phi_{\rm M}$ as
a function of their $F_{\rm M}$, $\langle F \rangle$, and $F_{1/2}$ is
also plotted in Fig.  \ref{fig:flatstat}.
The fraction of  oblate triaxial ellipsoids is rather  similar to that
of  the  complete sample,  being  $10\%$,  $78\%$,  and $83\%$  if  we
consider  bulges  with  $F_{\rm  M}<1$,  $\langle  F  \rangle<1$,  and
$F_{1/2}<1$, respectively.  The mean values of $\langle F \rangle$ and
$F_{1/2}$ are  0.88 and 0.71,  respectively, for the  complete sample,
and 0.86 and 0.75, respectively, for the selected sample.

\section{Intrinsic shape of bulges}
\label{sec:3d}

The  distributions   of  the  equatorial   ellipticity  and  intrinsic
flattening of bulges have  been studied in Sects.  \ref{sec:ellip} and
\ref{sec:flat} as  two independent and not  correlated statistics.  It
is possible to  find the relation between them  from Eqs.  \ref{eqn:E}
and \ref{eqn:Fdef}

\begin{equation}
\sqrt{E^2 - \sin^2{\phi_B}}=\frac{\frac{F}{F_{\rm \theta}}\sin{2\phi_C} + \sin{\phi_B} \cos{\left(2\phi_C - \phi_B\right)}}{\sin{\left(2\,\phi_C-\phi_B\right)}},
\label{eqn:shape1}
\end{equation}
%
to constrain the intrinsic shape of an observed bulge with the help of
the known  characteristic angles  $\phi_B$ and $\phi_C$,  which depend
only   on   the  measured   values   of   $a$,   $b$,  $\delta$,   and
$\theta$. Eq. \ref{eqn:shape1}  can be rewritten as a  function of the
axial ratios $B/A$ and $C/A$ as

\begin{eqnarray}
\frac{2\,\sin{\left(2\phi_C\right)}}{F_{\rm \theta}}\frac{C^2}{A^2} = 
\nonumber \\
\sin{\left(2\phi_C-\phi_B\right)} \sqrt{\left(1-\frac{B^2}{A^2}\right)^2 - \sin^2{\phi_B}\left(1+\frac{B^2}{A^2}\right)^2} -\nonumber \\
\sin{\phi_B}\cos{\left(2\phi_C - \phi_B\right)}\left(1+\frac{B^2}{A^2}\right)^2.
\label{eqn:vartie}
\end{eqnarray}

Since $B/A$ and $C/A$ are  both functions of the same variable $\phi$,
their probabilities are  equivalent (i.e., for a given  value of $B/A$
with probability  $P(B/A)$, the corresponding value  of $C/A$ obtained
by Eq. \ref{eqn:vartie} has a probability $P(C/A)=P(B/A)$).
This allows  us to obtain  the range of  possible values of  $B/A$ and
$C/A$  for  an observed  bulge  and  to  constrain its  most  probable
intrinsic  shape  by  adopting  the probabilities  $P(Z)$  and  $P(F)$
derived in Sects.  \ref{sec:ellip} and \ref{sec:flat}, respectively.

\begin{figure}[!h]
  \centering
  \includegraphics[width=0.5\textwidth]{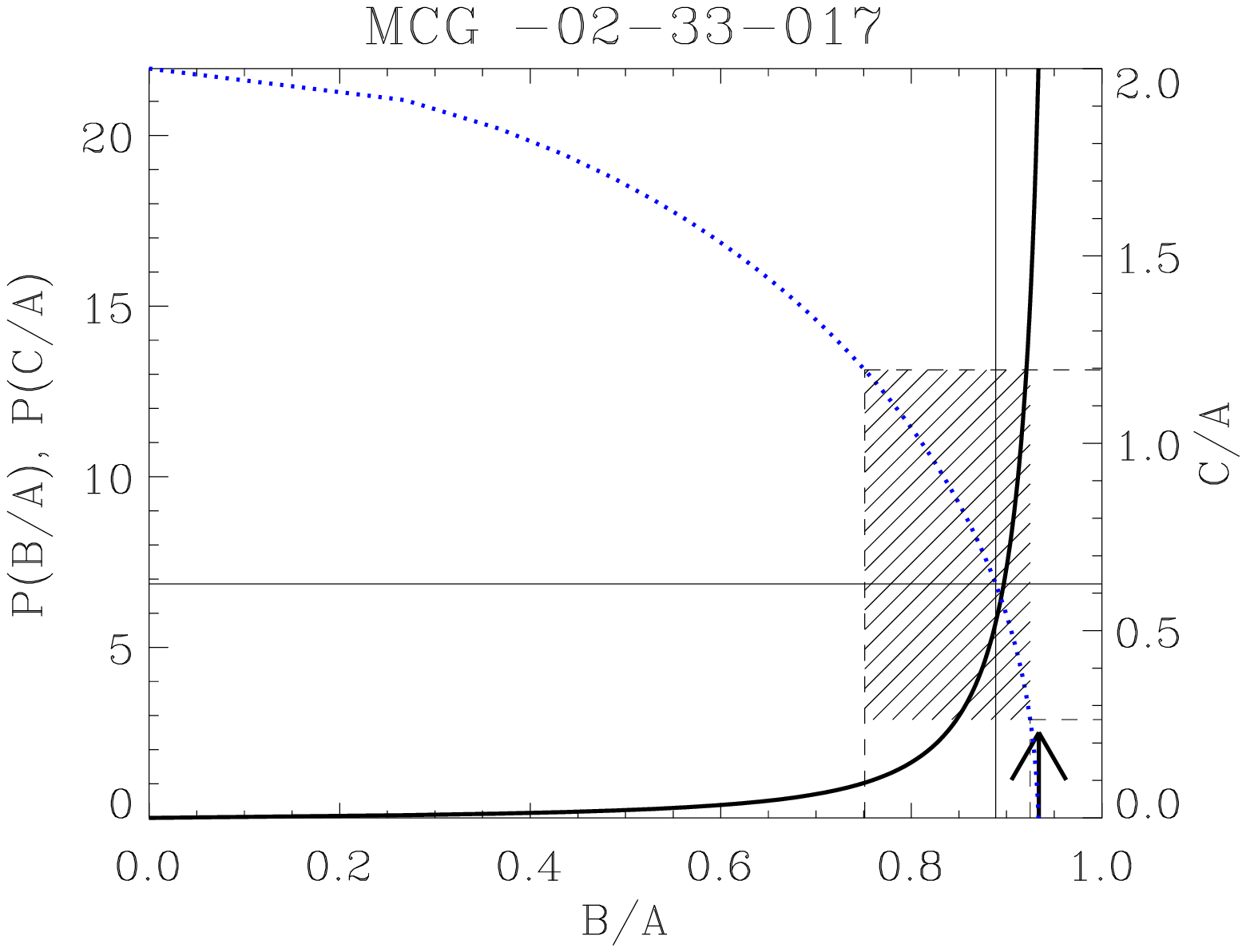}
  \includegraphics[width=0.5\textwidth]{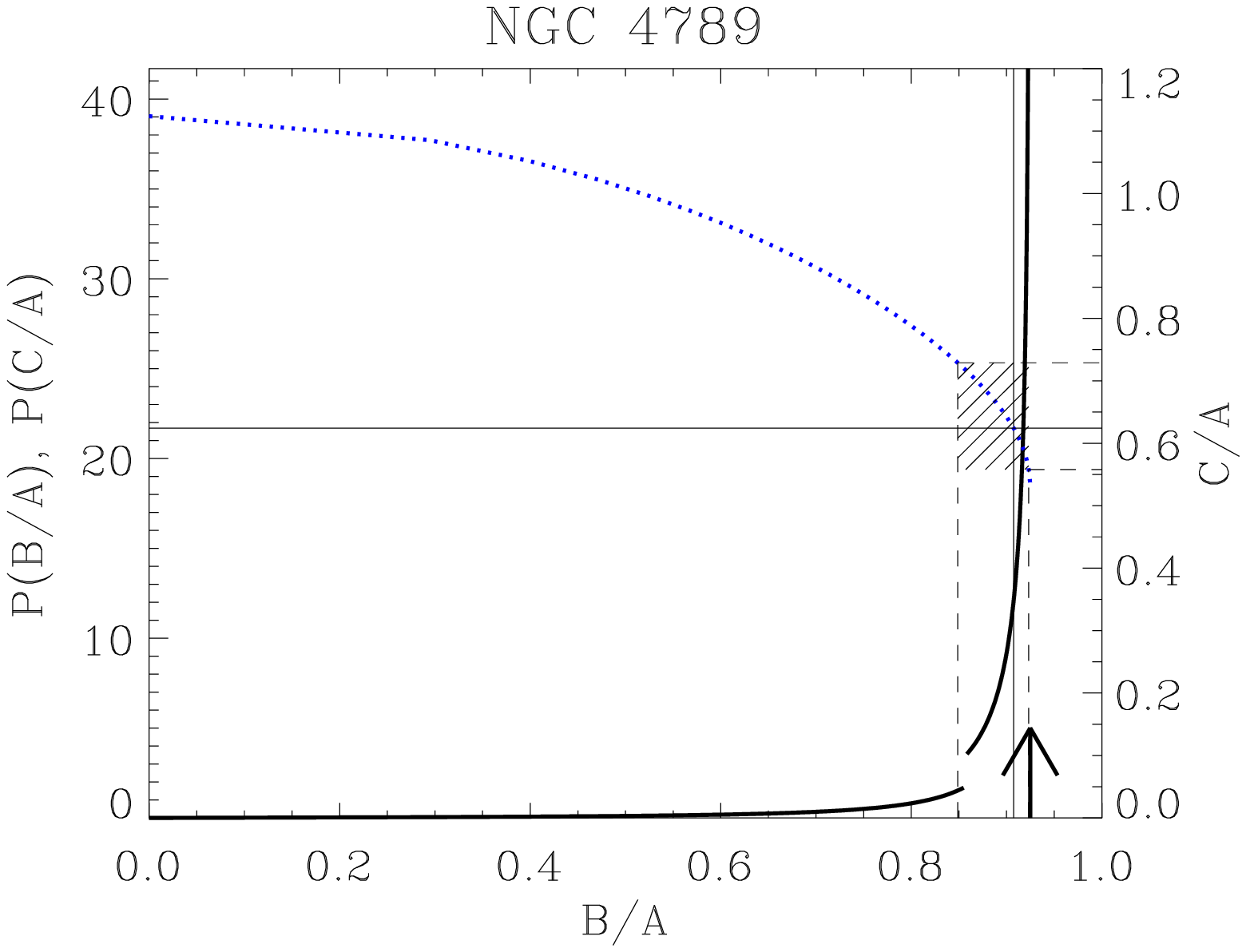}
  \caption{Relation between  the axial ratios $B/A$ and  $C/A$ for two
    sample  bulges. MCG~-02-33-017  (upper panel)  hosts a  bulge with
    $\phi_{M} < \phi_C$ and NGC~4789  (lower panel) hosts a bulge with
    $\phi_{M} >  \phi_C$. The probability associated  with each value
  of $B/A$  and its corresponding  value of $C/A$ (thick  solid line),
  the value of $C/A$ as a function of $B/A$ (dotted line), the maximum
  value of  the equatorial ellipticity  (arrow), the median  values of
  $B/A$ (vertical  thin solid line)  and $C/A$ (horizontal  thin solid
  line),  and the confidence  region which  encloses all  the possible
  values of $B/A$ and $C/A$ within a $67\%$ probability (hatched area)
  are shown in both panels.}
  \label{bavsca}%
  \end{figure}

An example of  the application of Eq.  \ref{eqn:vartie}  to two bulges
of our sample  is shown in Fig.  \ref{bavsca},  where the hatched area
marks  the  confidence  region  which  encloses $67\%$  of  the  total
probability  for all  the possible  values  of $B/A$  and $C/A$.   The
intrinsic  shape  of bulges  with  $\phi_C  <  \phi_{\rm M}$  is  less
constrained,  since the  median values  of  $B/A$ and  $C/A$ are  less
representative of their actual values.  This is the case for the bulge
of MCG~-02-33-017  (Fig. \ref{bavsca},  top panel).  On  the contrary,
the intrinsic shape  of bulges with $\phi_C >  \phi_{\rm M}$ is better
constrained.    This  is   the  case   for  the   bulge   of  NGC~4789
(Fig. \ref{bavsca}, bottom panel).

\subsection{Statistics of the intrinsic shape of bulges}
\label{sec:resanddis}

\begin{figure}[!h]
  \centering
  \includegraphics[width=0.5\textwidth]{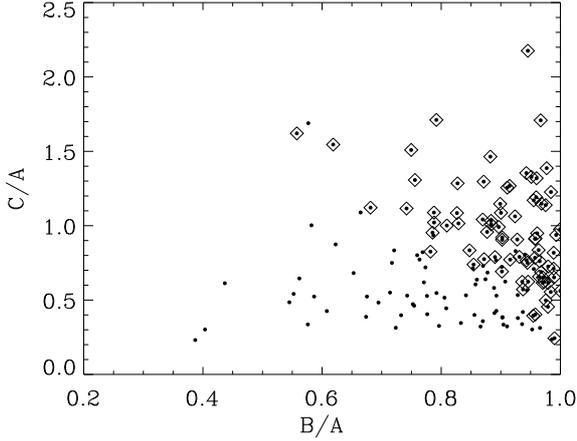}
  \caption{The intrinsic  shape of the  148 sample bulges.   The axial
    ratios  with  $50\%$  probability  are  plotted  for  each  bulge.
    Diamonds  refer to  the 115  sample bulges  with $\phi_C>\phi_{\rm
      M}$.}
  \label{bavsca_points}%
  \end{figure}

Following  the above  prescriptions,  we calculated  the axial  ratios
$B/A$  and $C/A$  and their  confidence intervals  for all  the sample
bulges.   There  is  no  correlation  between $B/A$  and  $C/A$  (Fig.
\ref{bavsca_points}), unless only bulges  with $\phi_C > \phi_{\rm M}$
are taken into account.  The  range of $C/A$ values corresponding to a
given  $B/A$ decreases  as  $B/A$ ranges  from  1.0 to  0.5, giving  a
triangular  shape  to  the   distribution  of  allowed  axial  ratios.
Circular and  nearly circular bulges  can have either  an axisymmetric
oblate  or  an  axisymmetric  prolate  or  a  spherical  shape.   More
elliptical bulges are more elongated along their polar axis.

We derived  the triaxiality parameter, as  defined by \citet{franx91},
for  the 115  sample bulges  with a  well-constrained  intrinsic shape
(i.e., those with $\phi_C>\phi_{\rm M}$)

\begin{equation}
T=\frac{1-\left(\frac{\hat{B}}{\hat{A}}\right)^2}
 {1-\left(\frac{\hat{C}}{\hat{A}}\right)^2} ,
\end{equation}
%
where $\hat{A}$, $\hat{B}$, and $\hat{C}$ are the lengths of the
longest, intermediate, and shortest semi-axes of the triaxial
ellipsoid, respectively (i.e., $\hat{A} \geq \hat{B} \geq \hat{C}$).
This  notation is different  with respect  to that  we adopted  in the
previous sections. Oblate triaxial (or axisymmetric) ellipsoids can be
flattened either  along the  $y-$axis on the  equatorial plane  of the
galaxy  or along the  polar axis.  Prolate triaxial  (or axisymmetric)
ellipsoids  can  be  elongated   either  along  the  $x-$axis  on  the
equatorial plane  of the  galaxy or along  the polar  axis. Therefore,
prolate bulges  do either lie  on the disk  plane (and are  similar to
bars) or do stick out from the disk (and are elongated perpendicularly
to it). This change of notation  is needed to compare our results with
those available in literature.

\begin{figure}[!h]
  \centering
  \includegraphics[width=0.5\textwidth]{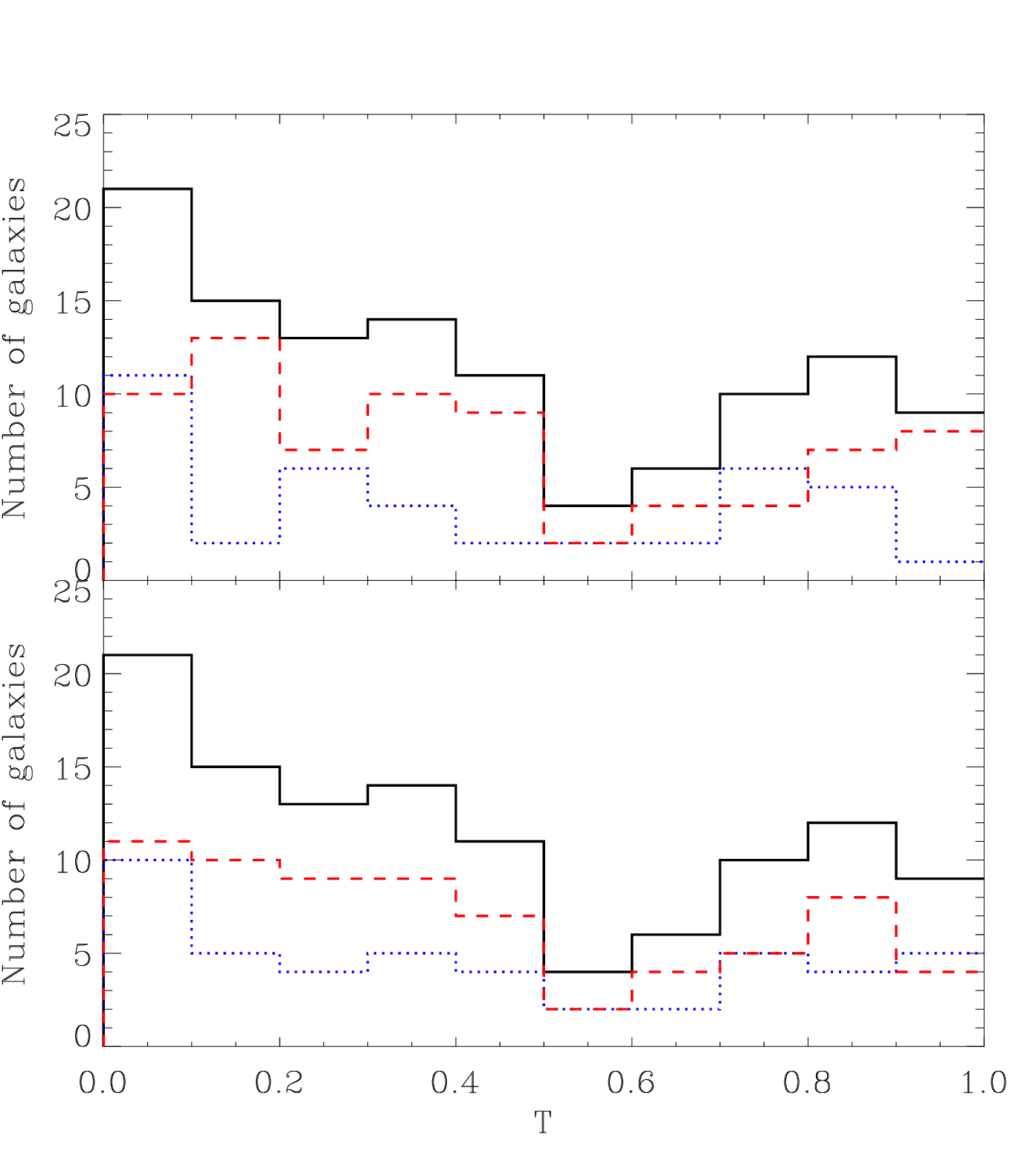}
  \caption{Distribution of the triaxiality parameter $T$ for the 115
    bulges with $\phi_C > \phi_{\rm M}$ (continuous line). The
    distributions of bulges with S\'ersic index $n\leq2$ (dotted line)
    and $n>2$ (dashed line) are shown in the upper panel.  The
    distributions of bulges of galaxies with bulge-to-total ratio
    $B/T\leq0.3$ (dotted line) and $B/T>0.3$ (dashed line) are shown
    in the lower panel.}
  \label{fig:T}
  \end{figure}

The triaxiality parameter  for bulges with $\phi_C >  \phi_{\rm M}$ is
characterized  by a  bimodal distribution  (Fig.  \ref{fig:T})  with a
minimum  at  $T=0.55$  and   two  maxima  at  $T=0.05$  and  $T=0.85$,
respectively.   According to  this distribution,  $65\%\pm4\%$  of the
selected  bulges  are  oblate  triaxial (or  axisymmetric)  ellipsoids
($T<0.55$)   and   the   remaining   $35\%\pm4\%$  are   prolate   (or
axisymmetric) triaxial ellipsoids ($T\ge0.55$).
The  uncertainties for  the  percentages were  estimated  by means  of
Montecarlo  simulations.   Since  $T$  is  a function  of  $\phi$,  we
generated  10000  random  values   of  $\phi$  in  the  range  between
$\phi_{B}$ and $\phi_{C}$ for each bulge and derived the corresponding
distributions of $B/A$  and $C/A$ according to their  PDFs. From $B/A$
and  $C/A$ we  calculated the  distribution  of $T$  and its  standard
deviation, which we adopted as uncertainty.

We  investigated the  cause of  such  a bimodality  by separating  the
bulges  according to  their  S\'ersic index  ($n$) and  bulge-to-total
luminosity  ratio  ($B/T$).  Both  quantities  were  derived for  each
sample  bulge in  Paper I.  The S\'ersic  index is  a  shape parameter
describing  the curvature  of  the surface-brightness  profile of  the
bulge. A profile with $n=1$ corresponds to an exponential law, while a
profile with $n=4$ corresponds to  an $r^{1/4}$ law. The bimodality is
driven by  bulges with S\'ersic index $n>2$  (Fig.  \ref{fig:T}, upper
panel), or  alternatively, by bulges of galaxies  with $B/T>0.3$ (Fig.
\ref{fig:T}, lower panel).
In fact, the sample of bulges with $\phi_C > \phi_{\rm M}$ and the two
subsamples  of  bulges with  $n>2$  and  of  bulges in  galaxies  with
$B/T>0.3$  are  characterized by  the  same  distribution  of $T$,  as
confirmed at  high confidence level ($>99\%$)  by a Kolmogorov-Smirnov
test.
We find  that $66\%\pm4\%$ of  bulges with $n>2$ have  $T<0.55$. Their
number  decreases as  $T$ increases  from  0 to  0.55.  The  remaining
bulges have  $T>0.55$ and  their number increases  as $T$  ranges from
0.55  to 1.  A  similar distribution  is  observed for  the bulges  of
galaxies  with $B/T>0.3$.   $67\%\pm4\%$  of them  host  a bulge  with
$T<0.55$.
Instead, the  distribution of the  triaxiality parameter of  bulges of
galaxies with $B/T\leq0.3$ is almost constant with a peak at $T=0.05$.
This is true  also for the bulges with $n\leq2$,  although to a lesser
degree.

The two subsamples of bulges with $n\leq2$ and $n>2$ are different, as
confirmed by  a Kolmogorov-Smirnov test ($99\%$  confidence level). In
particular,   the   fraction  of   oblate   axisymmetric  (or   nearly
axisymmetric)  bulges  ($T<0.1$)  is  remarkably higher  for  $n\leq2$
($27\%\pm4\%$)  than  for   $n>2$  ($14\%\pm3\%$).   The  fraction  of
triaxial  bulges  ($0.1  \leq  T  \leq 0.9$)  is  lower  for  $n\leq2$
($71\%\pm5\%$) than for $n>2$ ($76\%\pm3\%$).  The fraction of prolate
axisymmetric (or nearly axisymmetric) bulges ($T>0.9$) for $n\leq2$ is
$2\%\pm2\%$, but $11\%\pm3\%$ for $n>2$.

The two  subsamples of bulges  of galaxies with  $B/T > 0.3$  and $B/T
\leq 0.3$ are different too, as confirmed by a Kolmogorov-Smirnov test
($99\%$ confidence level).  The fraction of oblate axisymmetric bulges
($T < 0.1$)  is significantly higher for bulges  of galaxies with $B/T
\leq  0.3$ ($22\%\pm4\%$) than  for $B/T  > 0.3$  ($16\%\pm2\%$).  The
fraction of triaxial  bulges ($0.1 \leq T \leq  0.9$) is significantly
lower  for  $B/T  \leq  0.3$   ($67\%\pm4\%$)  than  for  $B/T  >  0.3$
($78\%\pm3\%$). A few prolate bulges ($T > 0.9$) are observed for $B/T
\leq 0.3$ ($11\%\pm3\%$) and $B/T > 0.3$ ($6\%\pm2\%$).
The distribution of  bulges with $n\leq2$ and bulges  of galaxies with
$B/T\leq0.3$  appears  to be  the  same  at  a high  confidence  level
($>99\%$) as confirmed by a Kolmogorov-Smirnov test.
  
Bulges with $\phi_C  > \phi_{\rm M}$ can be  divided into two classes:
those  with  $n\leq2$  (or  $B/T\leq0.3$)  and those  with  $n>2$  (or
$B/T>0.3$).   About  $70\%$ of  bulges  with  $n\leq2$  are hosted  by
galaxies with  $B/T\leq0.3$.  The same  is true for bulges  with $n>2$
which are mostly  hosted by galaxies with $B/T>0.3$.  This agrees with
the correlation between $n$ and $B/T$.

\begin{figure*}[!ht]
  \centering
  \includegraphics[width=12cm]{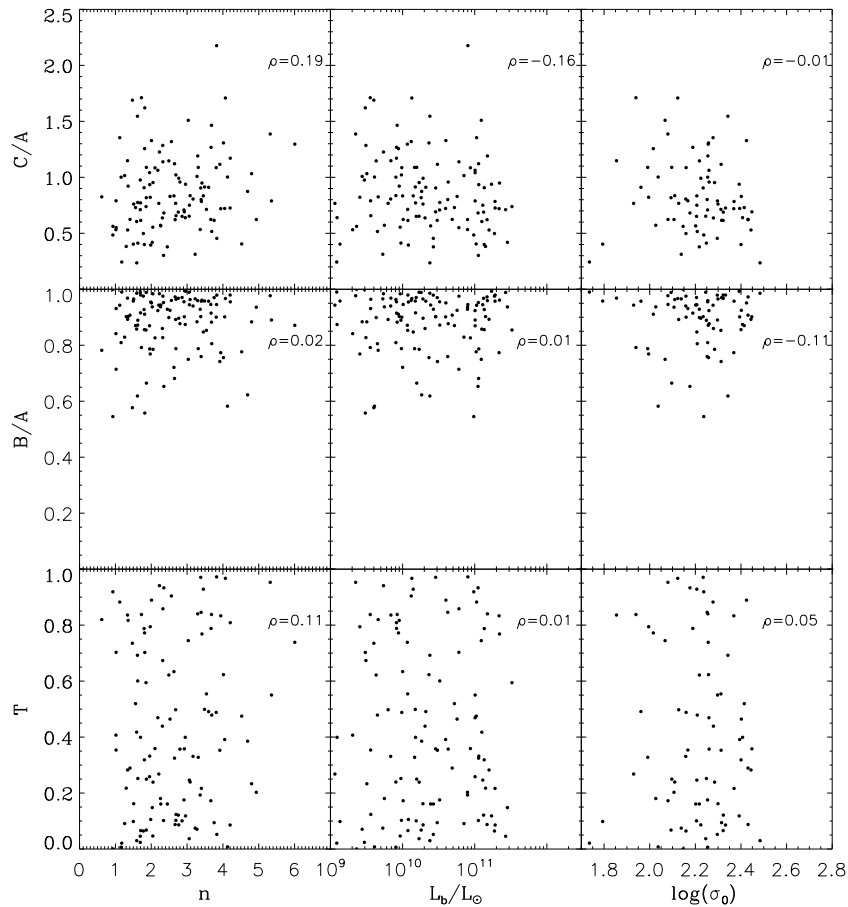}
  \caption{The bulge axial ratios  $C/A$ and $B/A$ and the triaxiality
    $T$ as  a function of  the bulge S\'ersic parameter  $n$, $J$-band
    luminosity $L_{\rm b}$ and central velocity dispersion $\sigma_0$.
    Only   the  115   bulges  with   $\phi_C  >   \phi_{\rm   M}$  are
    represented. The Spearman rank correlation coefficient ($\rho$) is
    shown in upper right corner of each panel.}
  \label{fig:correlations}
  \end{figure*}

In order to understand whether  the intrinsic shape is correlated with
some of  the bulge properties we  measured in Paper I,  we plotted the
axial ratios $C/A$ and $B/A$  and the triaxiality of the sample bulges
with $\phi_C  > \phi_{\rm M}$ as  a function of  their S\'ersic index,
$J$-band   luminosity,   and   central   velocity   dispersion   (Fig.
\ref{fig:correlations}).   As we found  in Paper  I for  the intrinsic
equatorial  ellipticity,   there  are  no   statistically  significant
correlations  between the bulge  shape and  the bulge  S\'ersic index,
luminosity or velocity  dispersion as pointed out by  the low Spearman
rank correlation  coefficient (Fig.  \ref{fig:correlations}). However,
this could be  a selection effect since the  sample of observed bulges
spans over a limited range of Hubble types (S0--Sb).

\subsection{The influence of nuclear bars on the intrinsic shape of bulges} 
 
Our sample  galaxies were selected  to not host large-scale  bars.  We
checked for their  presence in Paper I by a  visual inspection of both
the  original   image  and  the  residual  image   we  obtained  after
subtracting  the  best-fitting   photometric  model.   However,  these
selection  criteria did  not account  for the  presence  of unresolved
nuclear bars. Nuclear  bars are more elongated than  their host bulges
and  have  random  orientations,   therefore  they  could  affect  the
measurement of the structural parameters of bulges and consequently of
their intrinsic shape.

In Paper I we built up a set of 1000 artificial images with a S\'ersic
bulge, an  exponential disk,  and a Ferrers  nuclear bar to  study the
effects of the  bar on the measurements of  the photometric parameters
of  bulge and  disk. The  mean errors  on the  fitted axial  ratio and
position  angle of  the  bulge ($\langle  \Delta  q_{\rm b}  \rangle$,
$\langle \Delta  {\rm PA}_{\rm b} \rangle$) and  disk ($\langle \Delta
q_{\rm  d} \rangle$, $\langle  \Delta {\rm  PA}_{\rm d}  \rangle$) and
their standard  deviations ($\delta \Delta q_{\rm  b}$, $\delta \Delta
{\rm  PA}_{\rm b}$,  $\delta \Delta  q_{\rm d}$,  $\delta  \Delta {\rm
  PA}_{\rm d}$) are given in Table 2 of Paper I.

In  the present  paper,  we  tested whether  including  a nuclear  bar
affects the $T$ distribution.   For each galaxy, we randomly generated
a series  of 1000 values of  $q_{\rm b}$, $q_{\rm  d}$, ${\rm PA}_{\rm
  b}$, and  ${\rm PA}_{\rm  d}$. To assess  whether the  bulges appear
elongated and twisted with respect to  the disk due to the presence of
a  nuclear  bar,  we  assumed  that the  axial  ratios  were  normally
distributed  around the  values  $q_{\rm b}+\langle  \Delta q_{\rm  b}
\rangle$  and  $q_{\rm  d}+\langle  \Delta  q_{\rm  d}  \rangle$  with
standard  deviations  $\delta \Delta  q_{\rm  b}$  and $\delta  \Delta
q_{\rm d}$,  respectively, and that the position  angles were normally
distributed around the values  ${\rm PA}_{\rm b}\pm\langle \Delta {\rm
  PA}_{\rm b}  \rangle$ and ${\rm PA}_{\rm  d}\pm\langle \Delta q_{\rm
  d} \rangle$  with standard  deviations $\delta \Delta  {\rm PA}_{\rm
  b}$ and  $\delta \Delta {\rm  PA}_{\rm d}$, respectively.   We chose
the  PA values that  gave the  smallest $\delta$  with respect  to the
observed one.

\begin{figure} 
\centering 
\includegraphics[width=8.5cm]{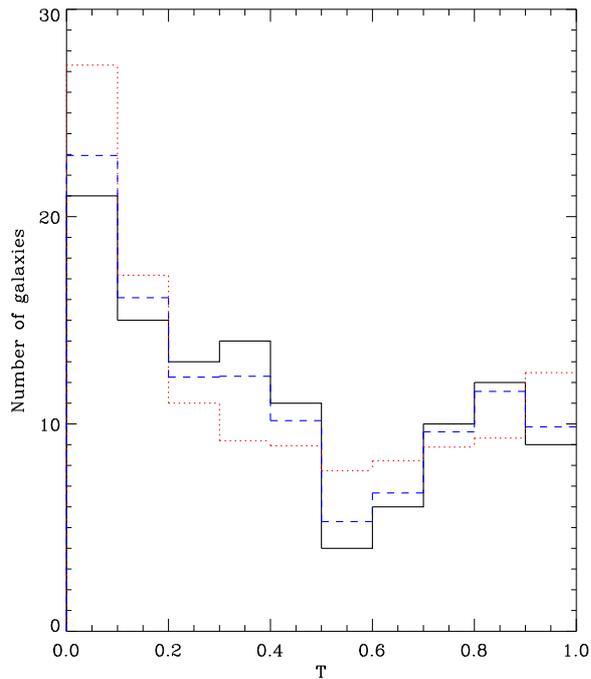} 
\caption{Distribution  of  the   triaxiality  parameter  $T$  for  the
  original  sample  of  115   bulges  with  $\phi_C  >  \phi_{\rm  M}$
  (continuous line), for a sample with $30\%$ of bulges with a nuclear
  bar (dashed line)  and for a $100\%$ fraction  of galaxies hosting a
  nuclear bar (dotted line).}
\label{fig:Tdistri_simu} 
\end{figure}

If we  assume that all  the artificial bulges  host a nuclear  bar, we
still    obtain    a     bimodal    distribution    of    $T$    (Fig.
\ref{fig:Tdistri_simu}). However, the  fraction of oblate axisymmetric
(or  nearly axisymmetric)  bulges ($T<0.1$)  is  higher ($23\%\pm2\%$)
with respect to the observed $18\%\pm3\%$.
For a  more realistic  fraction of galaxies  which host a  nuclear bar
(i.e, $30\%$, see \citealt{laine02}; \citealt{erwin04}), the resulting
distribution of $T$ is  consistent within errors with the distribution
derived  in Sect.  \ref{sec:resanddis}  (Fig. \ref{fig:Tdistri_simu}).
We  found  that $20\%\pm2\%$,  $71\%\pm3\%$,  and  $9\%\pm2\%$ of  the
sample bulges are oblate axisymmetric ($T < 0.1$), triaxial ($0.1 \leq
T \leq 0.9$), and prolate axisymmetric ($T > 0.9$), respectively, with
respect to the  $18\%\pm3\%$, $74\%\pm4\%$, and $8\%\pm2\%$ previously
found.
In  addition, we have also  tested the effects of  not consider a
  distribution of bar parameters but only the stronger bar included in
  the simulations  (0.8 $\times$ r$_{\rm  e}$, q$_{\rm b}$ =  0.2, and
  L$_{\rm  bar}$  =  0.02  $\times$  L$_{\rm tot}$),  i.e.,  the  {\it
    worst-case scenario} .   If we assume that 30$\%$  of our galaxies
  host  this  kind  of  nuclear  bars  the  results  change  strongly,
  obtaining that  only $56\%\pm4\%$ of the sample  bulges are triaxial
  ($0.1 \leq T \leq 0.9$)  with respect to the $74\%\pm4\%$ previously
  found.  Considering that all galaxies host such kind of nuclear bars
  the fraction of triaxial bulges is $30\%\pm3\%$.

The measured ellipticity  and bulge misalignment with the  disk of the
artificial galaxies without a nuclear  bar are smaller with respect to
the actual values  measured for the sample bulges.  This sets an upper
limit to the axisymmetry of the bulges.  
 
%
 
\section{Conclusions}
\label{sec:conclu}

In this work,  we have developed a new method  to derive the intrinsic
shape  of  bulges. It  is  based  upon  the geometrical  relationships
between  the  observed  and  intrinsic  shapes  of  bulges  and  their
surrounding disks.
We  assumed that  bulges  are triaxial  ellipsoids  with semi-axes  of
length $A$  and $B$ in  the equatorial plane  and $C$ along  the polar
axis. The  bulge shares the  same center and  polar axis of  its disk,
which is circular and lies on the equatorial plane of the bulge.
The intrinsic  shape of the  bulge is recovered from  photometric data
only. They include the lengths $a$  and $b$ of the two semi-major axes
of the ellipse, corresponding to the two-dimensional projection of the
bulge, the twist  angle $\delta$ between the bulge  major axis and the
galaxy line of nodes, and the galaxy inclination $\theta$.  The method
is completely independent of the  studied class of objects, and it can
be applied whenever a triaxial ellipsoid embedded in (or embedding) an
axisymmetric component is considered.

We  analyzed  the  magnitude-limited  sample of  148  unbarred  S0--Sb
galaxies,  for  which  we   have  derived  (Paper  I)  the  structural
parameters of bulges and disks by a detailed photometric decomposition
of their near-infrared surface-brightness distribution.

From  the study of  the equatorial  ellipticity $Z=B^2/A^2$,  we found
that there is a combination of the characteristic angles for which the
intrinsic shape can be  more confidently constrained.  This allowed us
to  select  a  qualified  subsample  of 115  galaxies  with  a  narrow
confidence interval  (corresponding to $67\%$ of  probability) of $Z$.
For example, bulges  with $B \approx A$ are  among those characterized
by the narrower confidence interval and the best determination of $Z$.
The fraction of selected  bulges with a maximum equatorial ellipticity
$Z_{\rm M} < 0.80$  ($B/A<0.89$), mean equatorial ellipticity $\langle
Z \rangle < 0.80$ and  a median equatorial ellipticity $Z_{1/2} < 0.80$
is $33\%$, $55\%$  and $43\%$, respectively. We conclude  that not all
the selected bulges have a  circular (or nearly circular) section, but
a  significant fraction  of  them is  characterized  by an  elliptical
section.  These bulges are strong  candidates to be triaxial. In spite
of the lower fraction of  bulges with a maximum equatorial ellipticity
smaller  than 0.8,  $Z_{\rm M}$  is a  good proxy  for  the equatorial
ellipticity because  the selected sample contains all  the bulges with
$B \approx A$.

The  analysis of the  intrinsic flattening  $F=2\,C^2/(A^2+B^2)$ shows
that only  a few  bulges of the  selected sample are  prolate triaxial
ellipsoids.  Only  $22\%$ and $17\%$ have a  mean intrinsic flattening
$\langle F  \rangle > 1$ or  a median intrinsic  flattening $F_{1/2} >
1$,  respectively.   The  fraction  rises  to $90\%$  when  a  maximum
intrinsic flattening $F_{\rm M} >  1$ is considered.  However, this is
due  to the  projection effect  of triaxial  ellipsoids.   Indeed, the
fraction of bulges  which are actually elongated along  the polar axis
is  very small: only  $18\%$ of  bulges with  $F_{\rm M}  > 1$  have a
probability greater than $50\%$ to have an intrinsic flattening $F>1$,
and there are no bulges with  more than a $90\%$ probability of having
$F>1$.   Thus, $F_{\rm  M}$  is not  a  good proxy  for the  intrinsic
flattening.

After considering the  equatorial ellipticity and intrinsic flattening
as independent parameters, we derived the relation among them in order
to calculate for each sample bulge both axial ratios, $B/A$ and $C/A$,
and their confidence intervals. As  already found for $Z$ and $F$, the
axial ratios  are better constrained  for the selected sample of
  115 bulges.
We derived  the triaxiality parameter, as  defined by \citet{franx91},
for all of them. We found that it follows a bimodal distribution with
a minimum  at $T=0.55$  and two maxima  at $T=0.05$  (corresponding to
oblate  axisymmetric or nearly  axisymmetric ellipsoids)  and $T=0.85$
(strongly prolate triaxial ellipsoids), respectively. According
to  this  distribution,  $65\%$  of  the selected  bulges  are  oblate
triaxial  (or axisymmetric)  ellipsoids ($T<0.55$)  and  the remaining
$35\%$ are prolate triaxial (or axisymmetric) ellipsoids ($T>0.55$).
This bimodality  is driven by  bulges with S\'ersic  index $n >  2$ or
alternatively by bulges of galaxies with a bulge-to-total ratio $B/T >
0.3$.  Bulges  with $n \leq 2$  and bulges of galaxies  with $B/T \leq
0.3$ follow  a similar distribution,  which is different from  that of
bulges with $n > 2$ and bulges of galaxies with $B/T > 0.3$.
In particular,  the sample of bulges  with $n\leq2$ and  the sample of
bulges  of galaxies  with $B/T  \leq 0.3$  show a  larger  fraction of
oblate  axisymmetric (or nearly  axisymmetric) bulges  ($T <  0.1$), a
smaller fraction of triaxial bulges ($0.1 \leq T \leq 0.9$), and fewer
prolate axisymmetric (or nearly  axisymmetric) bulges ($T > 0.9$) with
respect to  the corresponding sample  of bulges with  $n > 2$  and the
sample of bulges of galaxies with $B/T > 0.3$, respectively.

The different distribution of the intrinsic shapes of bulges according
to their S\'ersic  index gives further support to  the presence of two
bulge populations with  different structural properties: the classical
bulges,  which  are  characterized by  $n  >  2$  and are  similar  to
low-luminosity elliptical galaxies, and  pseudobulges, with $n \leq 2$
and   characterized   by   disk-like  properties   \citep[see][for   a
  review]{kormendykennicutt04}.
The correlation between the intrinsic  shape of bulges with $n \leq 2$
and those in galaxies with $B/T \leq 0.3$ and between bulges with $n >
2$ and those in galaxies with  $B/T > 0.3$ agrees with the correlation
between the bulge S\'ersic index  and bulge-to-total ratio of the host
galaxy,    as   recently    found    by   \citet{droryfisher07}    and
\citet{fisherdrory08}.

No statistically significant correlations  have been found between the
intrinsic   shape  of   bulges  and   bulge  luminosity   or  velocity
dispersion. However, this could  be a selection effect since the
sample bulges span a limited range of Hubble types (S0--Sb).

The observed bimodal distribution  of the triaxiality parameter can be
compared  to  the properties  predicted  by  numerical simulations  of
spheroid formation.
\citet{cox06} studied the structure of spheroidal remnants formed from
major  dissipationless  and dissipational  mergers  of disk  galaxies.
Dissipationless  remnants are  triaxial  with a  tendency  to be  more
prolate, whereas dissipational remnants  are triaxial and tend be much
closer to oblate.  This result  is consistent with previous studies of
dissipationless  and  dissipational mergers  \citep[e.g.,][]{barnes92,
  hernquist92, springel00, gonzalezgarciabalcells05}.
In addition,  \citet{hopkins10} used semi-empirical  models to predict
galaxy merger rates and contributions  to bulge growth as functions of
merger mass, redshift, and mass ratio. They found that high $B/T$
  systems tend  to form  in major mergers,  whereas low  $B/T$ systems
  tend to form from minor mergers.
In this framework, bulges with $n \leq 2$, which shows a high fraction
of oblate  axisymmetric (or nearly axisymmetric) shapes  and have $B/T
\leq 0.3$, could be the result of dissipational minor mergers.  A more
complex    scenario   including    both   major    dissipational   and
dissipationless  mergers  is  required   to  explain  the  variety  of
intrinsic shapes found for bulges with $n > 2$ and $B/T > 0.3$.

On the other hand, depending on the initial conditions \citep[see][and
  references  therein]{vietri90},   the  final  shape   of  the  early
protogalaxies  could  also   be  triaxial.   However,  high-resolution
numerical  simulations in  a cosmologically  motivated  framework that
resolves the bulge  structure are still lacking.  The  comparison of a
larger sample of  bulges with a measured intrinsic  shape and covering
the  entire Hubble sequence  with these  numerical experiments  is the
next logical step in addressing the issue of bulge formation.

\begin{acknowledgements}
We acknowledge  the anonymous referee for  his/her insightful comments
which  helped to  improve the  reading  and contents  of the  original
manuscript. JMA  is partially funded  by the Spanish MICINN  under the
Consolider-Ingenio  2010 Program  grant  CSD2006-00070: First  Science
with the GTC  (http://www.iac.es/consolider-ingenio-gtc). JMA and JALA
are  partially funded  by  the project  AYA2007-67965-C03-01.  EMC  is
supported by grant CPDR095001 by Padua University. ES acknowledges the
Instituto de  Astrof\'\i sica de  Canarias for hospitality  while this
paper was in progress.

\end{acknowledgements}

\bibliographystyle{aa} 
\bibliography{reference} 

\end{document}